\documentclass[acmtog, screen, nonacm]{acmart}
\usepackage{diagbox}
\usepackage{standalone}
\usepackage{tikz}
\usepackage{graphicx}
\usepackage{subcaption}
\usepackage{amsmath,bm}
\usepackage{booktabs} 
\usepackage{color}
\usepackage{soul}
\usepackage[normalem]{ulem}
\usepackage{xcolor}

\usepackage{wrapfig}
\usepackage[switch]{lineno}
\usepackage[boxed,ruled,linesnumbered]{algorithm2e} 

\SetCommentSty{mycommfont}
\let\oldnl\nl

\newcommand{\nonl}{\renewcommand{\nl}{\let\nl\oldnl}}
\newlength\savedwidth
\newcommand\whline[1]{\noalign{\global\savedwidth\arrayrulewidth
                               \global\arrayrulewidth #1} %
                      \hline
                      \noalign{\global\arrayrulewidth\savedwidth}}
                      
\usepackage{adjustbox}

\usepackage[ruled]{algorithm2e} 

\SetAlFnt{\small}
\SetAlCapFnt{\small}
\SetAlCapNameFnt{\small}
\SetAlCapHSkip{0pt}

\begin{document}
\title{Efficient GPU Cloth Simulation with Non-distance Barriers and Subspace Reuse}

\author{Lei Lan}
\orcid{0009-0002-7626-7580}
\affiliation{%
 \institution{University of Utah}
 \country{USA}}
\email{lanlei.virhum@gmail.com}

\author{Zixuan Lu}
\orcid{0000-0003-0067-0242}
\affiliation{%
 \institution{University of Utah}
 \country{USA}}
\email{birdpeople1984@gmail.com}

\author{Jingyi Long}
\orcid{0009-0007-2764-5123}
\affiliation{%
\institution{University of Utah}
\country{USA}}
\email{u6046121@umail.utah.edu}

\author{Chun Yuan}
\orcid{0009-0009-1134-0442}
\affiliation{%
 \institution{University of Utah}
 \country{USA}
}
\email{yuanchunisme@gmail.com}

\author{Xuan Li}
\orcid{0000-0003-0677-8369}
\affiliation{%
 \institution{UCLA}
 \country{USA}}
\email{xuan.shayne.li@gmail.com}

\author{Xiaowei He}
\orcid{0000-0002-8870-2482}
\affiliation{%
 \institution{Institute of Software, Chinese Academy of Sciences}
 \country{China}
}
\email{xiaowei@iscas.ac.cn}

\author{Huamin Wang}
\orcid{0000-0002-8153-2337}
\affiliation{%
 \institution{Style3D Research}
 \country{China}}
\email{wanghmin@gmail.com}

\author{Chenfanfu Jiang}
\orcid{0000-0003-3506-0583}
\affiliation{%
 \institution{UCLA}
 \country{USA}}
\email{chenfanfu.jiang@gmail.com}

\author{Yin Yang}
\orcid{0000-0001-7645-5931}
\affiliation{%
 \institution{University of Utah}
 \country{USA}}
\email{yangzzzy@gmail.com}

\begin{abstract}
This paper pushes the performance of cloth simulation, making the simulation interactive even for high-resolution garment models while keeping every triangle untangled. The penetration-free guarantee is inspired by the interior point method, which converts the inequality constraints to barrier potentials. We propose a major overhaul of this modality within the projective dynamics framework by leveraging an adaptive weighting mechanism inspired by barrier formulation. This approach does not depend on the distance between mesh primitives, but on the virtual life span of a collision event and thus keeps all the vertices within feasible region. Such a non-distance barrier model allows a new way to integrate collision resolution into the simulation pipeline. Another contributor to the performance boost comes from the subspace reuse strategy. This is based on the observation that low-frequency strain propagation is near orthogonal to the deformation induced by collisions or self-collisions, often of high frequency. Subspace reuse then takes care of low-frequency residuals, while high-frequency residuals can also be effectively smoothed by GPU-based iterative solvers. We show that our method outperforms existing fast cloth simulators by at least one order while producing high-quality animations of high-resolution models. 
\end{abstract}

%
%


%
%
\begin{teaserfigure}
\centering
\includegraphics[width=\textwidth]{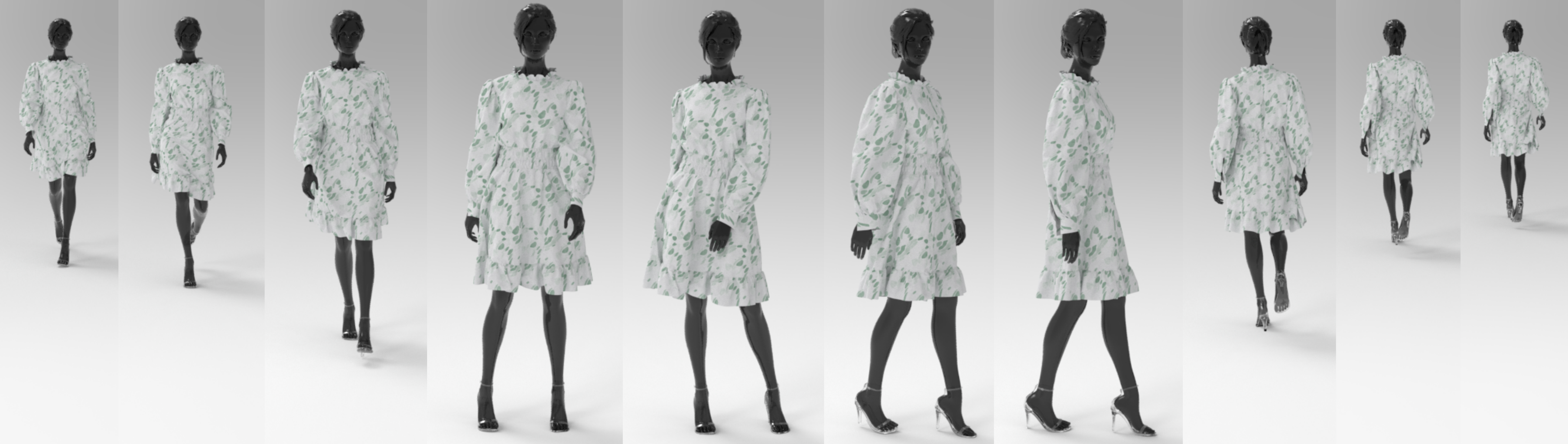}
\captionof{figure}{\textbf{Fashion show.}~~We present a new GPU-based cloth simulation framework with projective dynamics. Our method is able to simulate high-resolution cloth meshes at an interactive rate. With a non-distance-based barrier formulation, we can replace a large portion of traditional CCDs with the partial CCD procedure, which is much less expensive. The subspace reuse strategy relaxes the low-frequency errors effectively at the cost of single-digit milliseconds. Our method also features a residual forwarding trick to alleviate the damping issues generated by early termination and small-step line search filtering. In the teaser, we show an animated scene of the virtual fashion show. The model, dressed in a soft and light midi skirt, walks to the front and then turns around. These series of movements cause complex fabric dynamics, vividly showcasing the design concept of the garment. The garment is of high resolution and has $340$K vertices. The corresponding simulation involves over one million unknowns, and detailed local wrinkles can be well perceived. With a time step of $\Delta t = 1/200$ sec, the simulation runs at $4.8$ FPS. Please refer to the supplementary video for the corresponding animations.}
\label{fig:teaser}
\end{teaserfigure}


\maketitle
\section{Introduction}\label{sec:intro}
Cloth animation brings the simulated world to life in a vivid way, endows virtual characters with an infinite array of new appearances, and allows artists to lay their talents and inspirations on the triangular mesh. The primary challenge for today's cloth simulation arises from the irreconcilability between the desired visual quality and the limited computing resources -- it is often the case that the actual time budget allocated for the simulator is strictly capped e.g., in interactive design or games. Cloths and fabrics demonstrate intricate dynamics under collisions and contacts, yielding captivating fine deformations of wrinkles and folds. To faithfully capture those effects, a high-resolution mesh is preferred, and the increased number of degrees of freedom (DOFs) further stresses the simulation performance.  
This paper presents a GPU-based cloth simulation framework which is one or even two orders faster than the state-of-the-art GPU simulation algorithms for high-resolution cloth models.

Efficient processing of collision and self-collision of cloth is the pivotal concern for a high-performance cloth simulator. Our method is inspired by the recent success of incremental potential contact or IPC~\cite{li2020incremental,li2021codimensional}. When combined with CCD line search filtering and projection-Newton solver, IPC offers a non-penetration guarantee throughout the simulation process. It introduces a nonlinear repulsion between a pair of colliding or near-colliding primitives, which becomes infinitely strong if they get closer to each other. While this method has been proven robust, it requires repetitive CCDs (continuous collision detection) to calculate the distance between primitives in proximity -- this is costly for detailed meshes since all the triangles may be in contact. As a result, a dominant computation in many cloth simulators becomes the CCD processing. To address this issue, we design a novel barrier mechanism, which adaptively re-weights the collision constraints when the collision event exists over a continuous period of iterations, and does not depend on the actual distance between primitives. 
Because of this, most CCD procedures invoked in the simulation can be substantially simplified. 

On the solve side, we observe that existing GPU-based iterative solvers are less effective for smoothing low-frequency errors. In theory, this drawback could be remedied with model reduction or multigrid techniques, which project the system into a prescribed kinematic subspace. The difficulty comes from the nonlinearity of the system matrix, which varies under changing cloth poses, and constructing a subspace for each pose is not an option. To this end, we design a subspace reuse scheme that leverages a low-frequency rest-shape subspace for different deformed poses. This strategy well synergizes with the projective dynamics (PD) framework~\cite{bouaziz2014projective} because the geometric nonlinearity of cloth dynamics is taken care of in the local step, and the low-frequency subspace tends to be less sensitive to high-frequency deformations. We carefully exploit the structure of the global matrix in PD so that the expensive subspace projection of the full global matrix can be pre-computed. Our experiments show that subspace reuse cut the follow-up aggregated Jacobi iteration by $70\%$ on average. 

Our algorithm is also equipped with a residual forwarding scheme. As time-critical applications allocate a limited time budget for the simulator, early termination may lead to visual artifacts like over-stiffening and damping due to the CCD-based line search filtering. Residual forward estimates a ghost external force as the residual force inherited from the previous step to relieve this issue. 

In a nutshell, this paper proposes a simulator integrating several novel features to address the core challenges in cloth simulation. By deeply optimizing the pipeline, our method further pushes the quality and efficiency of the simulation. It is guaranteed that the resulting cloth poses are free of inter-penetration. Meanwhile, the combination of reused subspace and aggregated Jacobi iterations makes the solver effective for both low- and high-frequency deformations. More importantly, most calculations are friendly for GPU or any parallel computation platforms. The experiments show that our method delivers high-quality animation results while being one order faster than existing methods.

\section{Related Work}\label{sec:related}
Cloth simulation has been extensively studied in past decades. A vast volume of excellent contributions exists. Due to the page limit, this section only briefly surveys a few representative prior works.

\paragraph{Cloth simulation}
A common practice for cloth simulation is to discretize its geometry with a mass-spring network~\cite{choi2002stablecloth,liu2013massspring} or a triangle mesh~\cite{terzopoulos1987cloth,baraff1998cloth, volino2009cloth}. Early techniques use explicit integration with small time steps~\cite{provot1995deformation}. The stability is improved by switching to the implicit integration~\cite{baraff1998cloth}, at the cost of assembling and solving the resulting linearized systems. Cloth fabric is less extensible, showing strong resistance to stretching. The discrepancy between its stretching and bending behavior induces extra numerical difficulty. Strain limiting is a simple and effective approach to mitigate this challenge~\cite{provot1995deformation,thomaszewski2009strainlimiting,wang2010strainlimiting}. On the other hand, being unstretchable also inspires a simplified quadratic bending model~\cite{bergou2006qbend}. Cloth is more than an isotropic hyperelastic continuum. \citet{kim2020finite} reveals the underlying connection between the Baraff-Witkin model~\cite{baraff1998cloth} and anisotropic finite element method (FEM). One can finetune the strain-stress relation to obtain an accurate material model i.e., see~\cite{volino2009cloth}. Data-driven methods have also been used for this purpose~\cite{wang2011data,sperl2022estimation,feng2022learning}.

\paragraph{Collision processing}
Collision handling has always been an integral part of cloth simulation. Discrete collision detection (DCD) is an efficient method to identify the list of colliding primitives. As the name suggests, DCD detects the inter-penetration at a specific time instance, and it fails to capture all the collision events for models with thin geometries (such as garments) or fast-moving objects. On the other hand, CCD offers a more robust way to detect inter-penetration. Assuming the trajectory is linear within a time interval, CCD calculates the first time of impact (TOI) between two primitives. For triangularized surfaces, CCD is often modeled as the root finding of cubic equations. It is error-prone~\cite{wang2021CCD}, and robust numerical procedures are preferred~\cite{wang2014defending,Yuksel2022}. The brute-force collision detection among all the primitive pairs is prohibitive in general. A commonly adopted method is to use some bounding volume hierarchy (BVH)~\cite{langetepe2006geometric} to avoid excessive triangle-triangle intersection tests, a.k.a collision culling. Various BVH types have been explored such as AABB~\cite{bergen1997efficient}, OBB~\cite{gottschalk1996obbtree}, sphere~\cite{hubbard1995collision,james2004bd}, Boxtree~\cite{zachmann2002minimal}, spherical shell~\cite{krishnan1998rapid} and so on.

To resolve the intersected triangles, the penalty method has been a popular choice for its simplicity~\cite{guan2012drape}. \citet{buffet2019untangle} extend the implicit field of volumetric objects to open surfaces to resolve the inter-penetration of multiple-layer cloth. \citet{baraff2003untangling} and \citet{wicke2006untangle} employ an untangling method that applies repulsion forces to minimize the colliding region for cloth-cloth collisions. \citet{volino2006collision} and \citet{ ye2012contour} minimize the length of the collision contour of the colliding region. Instead of solving the collision and cloth dynamics in a two-way coupled manner, \citet{provot1997collision} and \citet{bridson2002collision} suggest a post-simulation step grouping penetrating vertices into impact zones. \citet{huh2001collision} decompose the impact zone into smaller colliding clusters based on their positions on the original mesh. \citet{harmon2008collision} relax impact zones with an inelastic projection, allowing relative tangential movement. 

Recently, an interior-point-based algorithm called incremental potential contact (IPC)~\cite{li2020incremental} has been proposed, which guarantees the simulation to be free of intersections. It has then been generalized to the simulation of cloth/thin shells~\cite{li2021codimensional}, rigid/stiff bodies~\cite{ferguson2021rigidipc,lan2022abd}, and curved meshes~\cite{ferguson2023high}. IPC is highly time-consuming as a CCD is needed at each nonlinear iteration to ensure the simulation results stay within the feasible region. \citet{lan2022penetration} approximate the logarithm barrier with an increasingly stronger quadratic function so that the simulation fits the PD framework~\cite{bouaziz2014projective}. \citet{lan2023second} decompose the global collision configuration into local stencils. To reduce the cost of CCD, \citet{wu2020fastrepulsion} use point-point distance constraints between triangle pairs to avoid intersections. Another relevant prior art is from \citet{ly2020projective}. This method is also based on PD, and it incorporates Signorini-Coulomb law~\cite{daviet2011hybrid,brogliato1999nonsmooth} using constraint projections in a semi-implicit way. With a matrix splitting scheme, it handles frictional contact robustly and efficiently. Our method is orthogonal to~\cite{ly2020projective} in the sense that we aim to improve the performance for contact resolution using a non-distance barrier-like formulation to adaptively adjust the collision weight instead of resorting to complementarity programming~\cite{moreau1988unilateral}. Our subspace reuse technique may also be used in~\cite{ly2020projective}.
\paragraph{GPU-based simulation}
In addition to collision, another computational bottleneck is the (nonlinear) system solve due to the use of implicit integration. A widely used strategy is to convert the force equilibrium (i.e., strong form) to the variational form (i.e., weak form)~\cite{variational2006,optimizational2015,decomposedOptimization2019}.  Doing so offers new perspectives to the simulation in the light of optimization such as constraint-based methods. One can locally and inexactly solve those constraints via the the constraint projection~\cite{goldenthal2007efficient}, which highlights the potential for parallelization. For instance, position-based dynamics~\cite{muller2007position,macklin2016xpbd} uses per-vertex constraint projection making the simulation matrix-free. Projective dynamics (PD)~\cite{bouaziz2014projective} presents a global and local alternation scheme to solve the nonlinear dynamic system. PD quickly became a popular simulation modality because its local projections are trivially parallelizable. Instead of solving its global system exactly e.g., using Cholesky factorization, iterative linear solvers can be used, such as Jacobi~\cite{wang2015chebyshev,lan2022penetration}, Gauss-Seidel~\cite{fratarcangeli2016vivace} and preconditional conjugate gradient (PCG)~\cite{tang2013gpu}, which bring significant speedups on the GPU. For more general and nonlinear models, sophisticated GPU algorithms are needed to decouple unknown DOFs to obtain the global solution~\cite{wangAndYang2016,tang2018,twoway2023,lan2023second}. 
\paragraph{Model reduction \& multigrid method}
Model reduction is an acceleration technique that reduces the simulation cost. It uses a set of reduced coordinates to pre-parameterize the simulation in the subspace and lowers the total number of unknown DOFs. Linear modal analysis offers the optimal subspace approximate around the rest shape~\cite{pentland1989modalanalysis,choi2005modalwarping,obrien2003modalanalysis}. Large and rotational deformations are less intuitive. For StVK models, one can pre-compute the coefficients of the reduced Hessian and the internal force~\cite{barbic2012modelreduction}. It is also possible to combine modal analysis with stiffness warping~\cite{muller2002stiffnesswarping} at per-vertex local frames~\cite{choi2005modalwarping}. However, doing so suffers from ghost forces when simulating free-floating objects. Another collection of contributions builds data-driven reduced models. For instance, \citet{kim2009skipping} use recent simulation results to construct the subspace at the simulation runtime. \citet{shen2021high} and \citet{fulton2019latent} use an autoencoder to encode the fullspace DOFs using the latent representation. 

Model reduction can also be coupled with PD. For instance, \citet{brandt2018hyper} design a reduced model for both local projection and global solve. It is highly efficient when the global matrix is constant (at the cost of missing some local deformations). When the global matrix varies due to collision and contact events or mesh's topology changes, efficiently estimating the updated subspace matrix becomes a challenge. A strategy is to use Woodbury formulation to leverage the pre-factorized global matrix for the rest shape and low-rank matrix update as in~\cite{modi2021emu,li2021interactive}. Apart from those existing methods, we employ a novel subspace reuse method -- the subspace constructed at the rest shape is used for relax low-frequency errors. This strategy requires minimum computational cost, while the high-frequency is left for GPU-based iterative solvers.

Multigrid method~\cite{trottenberg2001multigrid} is another common technique to boost simulation efficiency when a large number of DOFs is present. It was originally proposed to solve Poisson-like equations abounded in fluid simulation~\cite{mcadams2010multigrid, molemaker2008multigrid}. For deformable/cloth simulation, one needs to represent the dynamics at different levels. The geometric multigrid (GMG)~\cite{georgii2006deformablemultigrid} approaches this by generating spatial discretization (e.g., meshes or grids) of different resolutions. \citet{xian2019galerkin} further simplifies this process by sampling points from the finest grid to form coarser grids. The algebraic multigrid (AMG), on the other hand, approaches this by generating a subspace of the fine dynamics, which shares a similar nature of model reduction. For example, \citet{li2023subspace} utilize a B-spline subspace, and \citet{tamstorf2015samultigrid} built the subspace by QR decomposition on near-kernel components. \citet{wang2018clothmultigrid} integrate multigrid into a nonlinear optimization process, which updates the residue and system matrix periodically. Our method can be understood as a two-level multigrid for the global step solve. The reused subspace solve eliminates the low-frequency errors, leaving the high-frequency to the aggregated Jacobi iterations. 
\section{Background}\label{sec:back}
To make the exposition self-contained, we start with a brief review of projective dynamics and the distance-based interior point method for collision resolution. The reader can find more details from the relevant literature e.g.,~\cite{li2020incremental,lan2022penetration,bouaziz2014projective}. 

Given an implicit time integration scheme such as backward Euler, many state-of-the-art cloth simulators rely on the variational formulation of:
\begin{equation}\label{eq:var}
    \arg \min_{x} E =  I(x, \dot{x}) + \Psi(x), \; I =\frac{1}{2h^2}\|\mathsf{M}^{\frac{1}{2}}(x - z)\|^2.
\end{equation}
Here, $x$ is the unknown variable we would like to compute at the next time step i.e., the position of all the cloth vertices. We also have:
\begin{equation}\label{eq:z}
z = x^* + h \dot{x}^* + h^2 \mathsf{M}^{-1} f_{ext},   
\end{equation}
as a known vector based on the previous position $x^*$, velocity $\dot{x}^*$, and an external force $f_{ext}$. $\mathsf{M}$ is the mass matrix, and $h$ is the time step size. The objective function $E$ consists of the inertia momentum ($I$) offering a mass-weighted regularization over $x$, and the elasticity potential ($\Psi$) controlling the deformation of the cloth. 

Under the framework of PD, Eq.~\eqref{eq:var} is split into two stages in the form of local-global (LG) iterations. In the local stage, the unknown DOFs are duplicated at individual constraints, which measure the strain/deformation under various metrics like the change of the edge length or the bending angle. The local step is formatted as:
\begin{equation}\label{eq:local}
    \arg\min_{y_i}\frac{1}{2}\|\mathsf{A}_i \mathsf{S}_i x - \mathsf{B}_i y_i \|^2,\; \text{s.t.}\, C_i(y_i) = 0. 
\end{equation}
In other words, the local step computes a \emph{target position} $y_i$, which not only satisfies the constraint $C_i$ exactly but is also closest to the current value of $x_i$ i.e., a projection-like operator. Here, $\mathsf{S}_i$ is a selection matrix picking DOFs relevant to $C_i$ from $x$ such that $x_i = \mathsf{S}_i x$. $\mathsf{A}_i$ and $\mathsf{B}_i$ map the positional information of $x_i$ and $y_i$ to the specific coordinate that the constraint $C_i$ measures. For instance, they can be a differential operator computing the deformation gradient of a triangle. The local step is highly parallelizable as the computation at each constraint is independent. 

The global stage follows as a standard linear solve:
\begin{equation}\label{eq:global}
    \left(\frac{\mathsf{M}}{h^2}+\sum_i w_i\mathsf{S}_i^\top\mathsf{A}_i^\top\mathsf{A}_i\mathsf{S}_i\right)x = \frac{\mathsf{M}}{h^2} z + \sum_i w_i\mathsf{S}_i^\top\mathsf{A}_i^\top\mathsf{B}_i y_i.
\end{equation}
Intuitively, the goal of Eq.~\eqref{eq:global} is to blend duplicated DOFs $y_i$ to produce a global solution of $x$ since $x_i$ could have multiple replicates if it is involved in several constraints. As a result, the weight, i.e., $w_i$ in Eq.~\eqref{eq:global} embodies the \emph{priority} of a constraint -- a bigger $w_i$ (relative to other constraints) means the global solve favors $x$ being closer to the corresponding $y_i$. In an extreme case when $w_i\rightarrow \infty$, $x_i \rightarrow y_i$ ignoring all the other constraints. One should not confuse $w_i$ with the \emph{stiffness} of the constraint: a high constraint stiffness produces a big internal force, which could overshoot and must be coupled with a line search for extra safeguards.

The presence of the collision and self-collision introduces a new energy into Eq.~\eqref{eq:var}:
\begin{equation}\label{eq:var_barrier}
    \arg \min_x E = I(x, \dot{x}) + \Psi(x) + B(x).
\end{equation}
Codimensional geometries of clothes make the simulation sensitive to inter-penetrations -- once collisions or self-collisions are generated, the fabric often gets more and more tangled in the following time steps. 
IPC~\cite{li2020incremental} offers a potential solution to the challenge, which formulates $B(x)$ as a log-barrier such that:
\begin{equation}\label{eq:ipc}
B_i =
\left\{
\begin{array}{ll}
\displaystyle -\kappa (d_i - \hat{d})^2\ln \left(\frac{d}{\hat{d}}\right), & 0 < d_i < \hat{d}\\
\displaystyle 0, & d_i \geq \hat{d}.
\end{array}
\right.
\end{equation}
Here, $\hat{d}$ is a user-provided tolerance of the collision resolution. $d_i(x)$ denotes the closest distance between the $i$-th pair of surface primitives, either a vertex-triangle pair or an edge-edge pair. $B(d_i)$ diverges if $d_i < \hat{d}$ and approaches $\infty$ when $d_i \rightarrow 0$. Consequently, as long as we keep $d_i$ positive at the beginning of a time step (i.e., all primitives are separate), the existence of $B(d_i)$ prevents any future inter-penetration with an increasingly stronger repulsion. \citet{lan2022penetration} further showed that IPC barrier function can also be integrated into LG iterations by setting the weighting function $w_i$ of the collision constraint as $B(d_i)$.

\section{Non-distance Barrier}\label{sec:exp}
While IPC \cite{li2021codimensional} and its PD variations~\cite{lan2022penetration} offer robust treatment for collisions, they are all based on $d_i$. Here, we name this family of barrier functions as the distance-based barrier or DBB. Updating DBB is expensive -- one needs to perform a broad phrase collision culling to generate the list of vertex-triangle or edge-edge pairs, compute TOI for each pair, identify the smallest TOI which is between zero and one of all pairs, compute $d_i$, and eventually obtain the latest value of $\sum B(d_i)$. Such a \emph{full CCD procedure} is invoked frequently during LG iterations~\cite{lan2022penetration} and becomes the dominant computation along the pipeline. 

We argue that not all DBBs are indispensable, and most of them can be replaced with a more economic alternative model under the PD framework. The underlying philosophy of DBB is to offer a nonlinear penalty, which becomes stiffer when the collision constraint is about to be violated. DBB geometrically approximates the indicator functions ($\delta_{\mathcal{A}}(x) = 0$, if $x \in \mathcal{A}$; $\delta_{\mathcal{A}}(x) = \infty$, if $x \notin \mathcal{A}$). A DBB itself is \emph{not} physically accurate i.e., the gradient of DBB differs from the actual collision force unless $\kappa$ in Eq.~\eqref{eq:ipc} is close to zero (i.e., the complementary slackness is sufficiently satisfied). In practice, what we want is an increasingly strong repulsion to correct the constraint violation, and this goal can be enabled without $d_i$, the actual distance between a pair of primitives. To this end, we design a non-distance barrier weight with exponential formulation:
\begin{equation}\label{eq:exp}
    w_i = B_i^{NDB} = k{K^{a_i}},
\end{equation}
where $a_i\in\mathbb{Z}^+$ represents how many \emph{consecutive LG iterations} has the constraint $C_i$ been active.
A contact constraint is considered active if the primitive pair remains sufficiently close to each other. $k$ is an initial weight, and $K$ is the base of the growth rate.  We use NDB to denote this simplified barrier model. Similar to DBB, NDB approaches to $\infty$ if a collision constraint $C_i$ remains unresolved over several LG iterations.
In this case, the local target position of the constraint shall be satisfied with the highest priority in the global solve. Unlike DBB on the other hand, NDB no longer depends on the distance between primitives or any physically/geometrically meaningful measures. It becomes a self-adjusting variable as the optimization proceeds with minimum computation costs. As we elaborate in the next subsection, this feature delivers substantial convenience and speedup for cloth simulation.
\paragraph{Discussion} DBB takes the primitive distance $d(x)$ as its parameter. In contrast, our non-distance barrier (NDB) strategy does not explicitly depend on the collision distance. Instead, it employs the collision's life span (i.e., number of iterations) to adaptively increase the weight of the constraint. NDB is not a barrier function in a strict sense since NDB is not explicitly related to $x$. 
Nevertheless, it introduces a penalty that asymptotically tends to infinity to prevent the interpenetration between primitives. This feature ensures the optimization variable $x$ remains within the feasible region, and works in a similar way as the conventional interior-point method. One limitation, however, is the accuracy of such a weighting mechanism. Since the iteration index is a discrete count, it is sometimes challenging to produce sufficiently different repulsion forces to distinguish nearby vertices. More iterations are therefore needed.

\subsection{Partial CCD}
An immediate advantage of using NDB is the alleviation of computational effort required for CCD. When updating $B_i^{NDB}$, it is sufficient to determine whether a collision remains active after the previous LG iteration, requiring a true-or-false response rather than an exact time of impact. This characteristic enables a more streamlined approach for NDB CCD, transforming the cubic root-finding problem into a series of dot product calculations, a technique we refer to as partial CCD.



Starting from a collision-free configuration, if two primitives $\mathcal{P}_1$, $\mathcal{P}_2$ collide with each other within a normalized time interval $(0, 1]$, there exist two points $p_1 \in \mathcal{P}_1$ and $p_2 \in \mathcal{P}_2$ which have intersecting trajectories such that:
\begin{equation}\label{eq:intersect}
    p^0_1 + t^*\left(p_1^1 - p^0_1\right) = p^0_2 + t^*\left(p_2^1 - p^0_2\right),
\end{equation}
where $p^0_{1,2}$ and $p_{1,2}^1$ denote the positions of those two points at the beginning and end of the time interval. $t^* \in (0, 1]$ is the intersection time.
With some manipulations, Eq.~\eqref{eq:intersect} can be re-written as:
\begin{equation}\label{eq:collision_judge}
    \left(p_2^1 - p_1^1\right) \cdot \left(p^0_2 - p^0_1 \right) 
    = \frac{t^* - 1}{t^*}\left(p^0_2 - p^0_1\right) \cdot \left(p^0_2 - p^0_1\right) \le 0.
\end{equation}
Eq.~\eqref{eq:collision_judge} suggests a non-positive inner product of $\big(p_2^1 - p_1^1\big) \cdot \big(p^0_2 - p^0_1 \big)$ at certain locations on $\mathcal{P}_1$ and $\mathcal{P}_2$ being a necessary condition of the collision between primitives.

We then build a query function based on l.h.s. of Eq.~\eqref{eq:collision_judge} by parameterizing $p_1$ and $p_2$ with $\lambda = [\lambda_1, \lambda_2]^\top$ in their corresponding primitives:
\begin{equation}\label{eq:query}
    Q(\lambda) = \left(p_2^1(\lambda) - p_1^1(\lambda)\right) \cdot \left(p^0_2(\lambda) - p^0_1(\lambda)\right).
\end{equation}
\setlength{\columnsep}{5 pt}

For a vertex-triangle pair specified by $x$, $x_1$, $x_2$, $x_3$ i.e., $x$ is the position of the vertex and $x_{1,2,3}$ are the three vertices of the triangle as shown in the inset on right, we have $p_1 = x$ and $p_2(\lambda) = x_1 + \lambda_1 (x_2 - x_1) + \lambda_2 (x_3 - x_1)$, where $\lambda_1$, $\lambda_2$ and $1 - \lambda_1 - \lambda_2$ are barycentric coordinates of $p_2$ in the triangle. Similarly, for an edge-edge pair specified by $x_{1,2}$ and  $x_{3,4}$, $p_{1,2}$ are: $p_1 = x_{1} + \lambda_1 (x_2 - x_1)$, $p_2 = x_3 + \lambda_2 (x_4 - x_3)$, for $0 \leq \lambda_1, \lambda_2 \leq 1$. Let $\Omega_\lambda$ be the domain of the query function $Q$. It is easy to see that $\Omega_\lambda$ forms a triangle $\lambda_{1,2} \geq 0, \lambda_1 + \lambda_2 \leq 1$ for a vertex-triangle pair, and a box $\lambda_{1,2}\in [0, 1]$ for an edge-edge pair. 

\begin{wrapfigure}{r}{0.3\linewidth}
    \includegraphics[width=\linewidth]{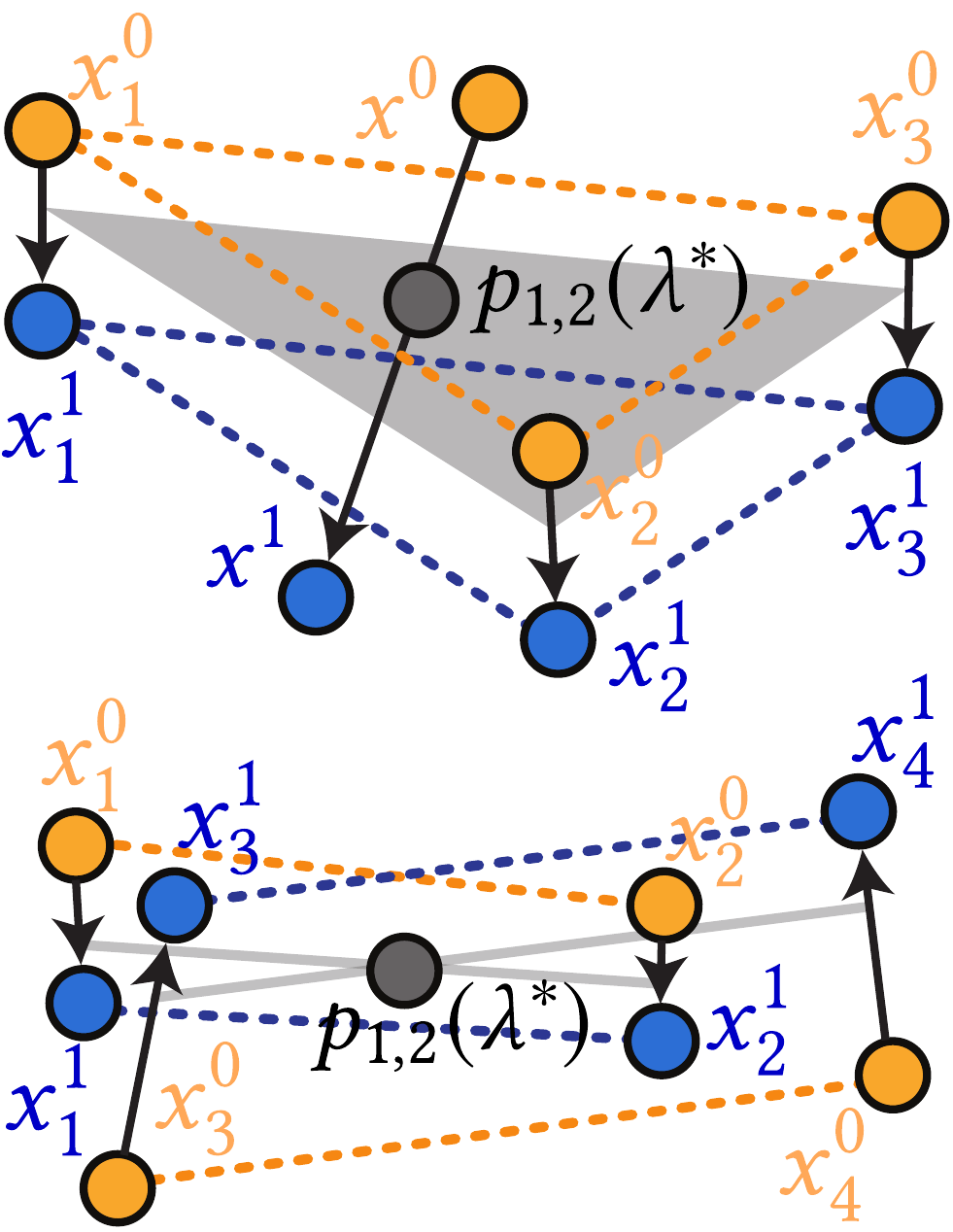}
\end{wrapfigure}

Since $Q(\lambda)$ is continuous, $Q < 0$ prescribes some neighborhoods around $\lambda^*$, where $p_1(\lambda^*)$ and $p_2(\lambda^*)$ converges at $t = t^*$. Instead of solving $\lambda^*$ and the corresponding $t^*$ i.e., as in most CCD algorithms, we query the value of $Q(\lambda)$ at multiple sample points in $\Omega_\lambda$. As long as our sampling is sufficiently dense to not miss those neighborhoods of $Q<0$, and all the queried $Q(\lambda)$ values are positive, we conclude that no collision occurs between $\mathcal{P}_1$ and $\mathcal{P}_2$. Eq.~\eqref{eq:query} does not involve $t^*$, meaning we skip the calculation for the actual TOI and only compute inner products of 3-vectors. The question is: how dense should the sampling be? 

To answer this question, we first set up the metric of sampling density. Let $\mathcal{S} = \left\{ \lambda_{(0)}, \lambda_{(1)},...\right\} \subset \Omega_\lambda$ be the set of sample points. We define that local sample interval $\rho(\lambda)$ for any $\lambda\in\Omega_\lambda$, $\lambda \neq \lambda_{(i)}$ is the distance from $\lambda$ to its nearest sample point in the parameter space. More formally, $\rho(\lambda)$ gives the largest radius of the disc $\Omega(\lambda)$ centered at $\lambda$ such that $\Omega(\lambda) \cap \mathcal{S} = \emptyset$. If $\lambda = \lambda_{(i)}$ happens to be a sample point, its local sample interval is zero. The \emph{sample interval} of the whole sample set is defined as: $\rho_\mathcal{S}= \max_{\lambda \in \Omega_\lambda} \rho(\lambda)$. 
Based on Lagrange Remainder Theorem, there exists an upper bound of $Q(\lambda)$ for any $\lambda \in \Omega_\lambda$:
\begin{equation}
    Q(\lambda) \le Q(\lambda^*) + \left(\max_{\lambda \in \Omega_\lambda} \left\| \nabla_\lambda Q \right\| \right) \left \| \lambda - \lambda^* \right\|,
\end{equation}
which leads to a suffcient condition of $Q(\lambda) \leq 0$:
\begin{multline}
     Q(\lambda^*) + \left(\max_{\lambda \in \Omega_\lambda} \left\| \nabla_\lambda Q \right\| \right) \left \| \lambda - \lambda^* \right\| \leq 0 \Rightarrow \\
       \left \| \lambda - \lambda^* \right\| \leq \rho = -\frac{Q(\lambda^*)}{\displaystyle \max_{\lambda \in \Omega_\lambda} \left\| \nabla_\lambda Q \right\|}.
\end{multline}

As the cloth is less stretchable, we assume that edge lengths of $\mathcal{P}_1$ and $\mathcal{P}_2$ are bounded by $L$. Given a finite velocity, the distance between $\mathcal{P}_1$ and $\mathcal{P}_2$ is bounded above by $H^1$ during $t\in (0, 1]$, and bounded below by $H^0$ at $t = 0$.
We can then obtain an upper bound of the norm of $\nabla_\lambda Q$:
\begin{equation}\label{eq:derivative_bound}
    \left| \frac{\partial Q}{\partial \lambda_{1, 2}} \right | \le 2 L (H^1 + 2L) \; \Rightarrow\;  \left \| \nabla_\lambda Q \right\| \le 2\sqrt{2} L (H^1 + 2L).
\end{equation}

According to Eq.~\eqref{eq:collision_judge}, we also have:
\begin{equation}
Q(\lambda^*) = \frac{t^* - 1}{t^*}\left\|p^0_2 - p^0_1\right\|^2 \le \frac{t^* - 1}{t^*}{(H^0)}^2.
\end{equation}
Together with Eq.~\eqref{eq:derivative_bound}, a thresholding sample interval over $\Omega_\lambda$ can then be obtained:
\begin{equation}\label{eq:sample_bound}
    \rho^* = \left(\frac{1}{\alpha} - 1\right)\frac{\left(H^0\right)^2}{2\sqrt{2} L (H^1 + 2L)}.
\end{equation}
When $\rho_{\mathcal{S}}$ is smaller than $\rho^*$, there is at least one sample point sitting in a neighborhood of $Q < 0$. Here, $\alpha < 1$ is the hyperparameter used in line search filtering, which is typically set as $0.8$.

\paragraph{Discussion}
Partial CCD is sample-based and can be considered as an \emph{approximation} of exact CCD algorithms~\cite{brochu2012efficient}. We would like to mention that our method does not exclusively rely on partial CCD. In other words, partial CCD, together with exponential-based NDB collision constraint projection works as a warm start, and regular CCDs are still performed during the simulation (but at a much lower frequency). Therefore, we do not need to set the sampling density to $\rho^*$ in practice (as this theoretical bound depends on various factors and varies at each iteration). The performance and the quality of the simulation are not sensitive to partial CCD accuracy (e.g., see Fig.~\ref{fig:exp_dbb}).

Eq.~\eqref{eq:sample_bound} appears sensitive to $H^0$, the closest distance between $\mathcal{P}_1$ and $\mathcal{P}_2$ for $t \in (0, 1]$. For instance, if a vertex is in the proximity of the triangle at $t = 0$ and hits the triangle at $t = t^*$, $\rho(\lambda^*)$ of the corresponding contact approaches zero i.e., the disc of $\Omega(\lambda^*)$ shrinks to a point. This issue can be easily fixed by setting the projection of the vertex on the triangle as a sample point if they are close to each other. In some extreme cases, where the cloth is substantially stretched or accelerated, we increase the sample density by scaling the sample interval by the change ratio of $\rho^*$.

A quick benchmark well demonstrates the potential of such a sample-based strategy: solving a cubic equation on the GPU as in the original IPC implementation~\cite{li2020incremental} is slower than evaluating Eq.~\eqref{eq:collision_judge} on a \texttt{RTX} \texttt{3090} GPU by over $6,000$ times. Partial CCD is about $2,000$ times faster than the state-of-the-art polynomial solver~\cite{Yuksel2022}. For a high-resolution cloth model, we do not need a lot of sample points on a primitive, making partial CCD faster than regular CCD by orders of magnitude.

\section{Subspace Reuse}\label{sec:reuse}
\begin{figure}
  \includegraphics[width=0.8\linewidth]{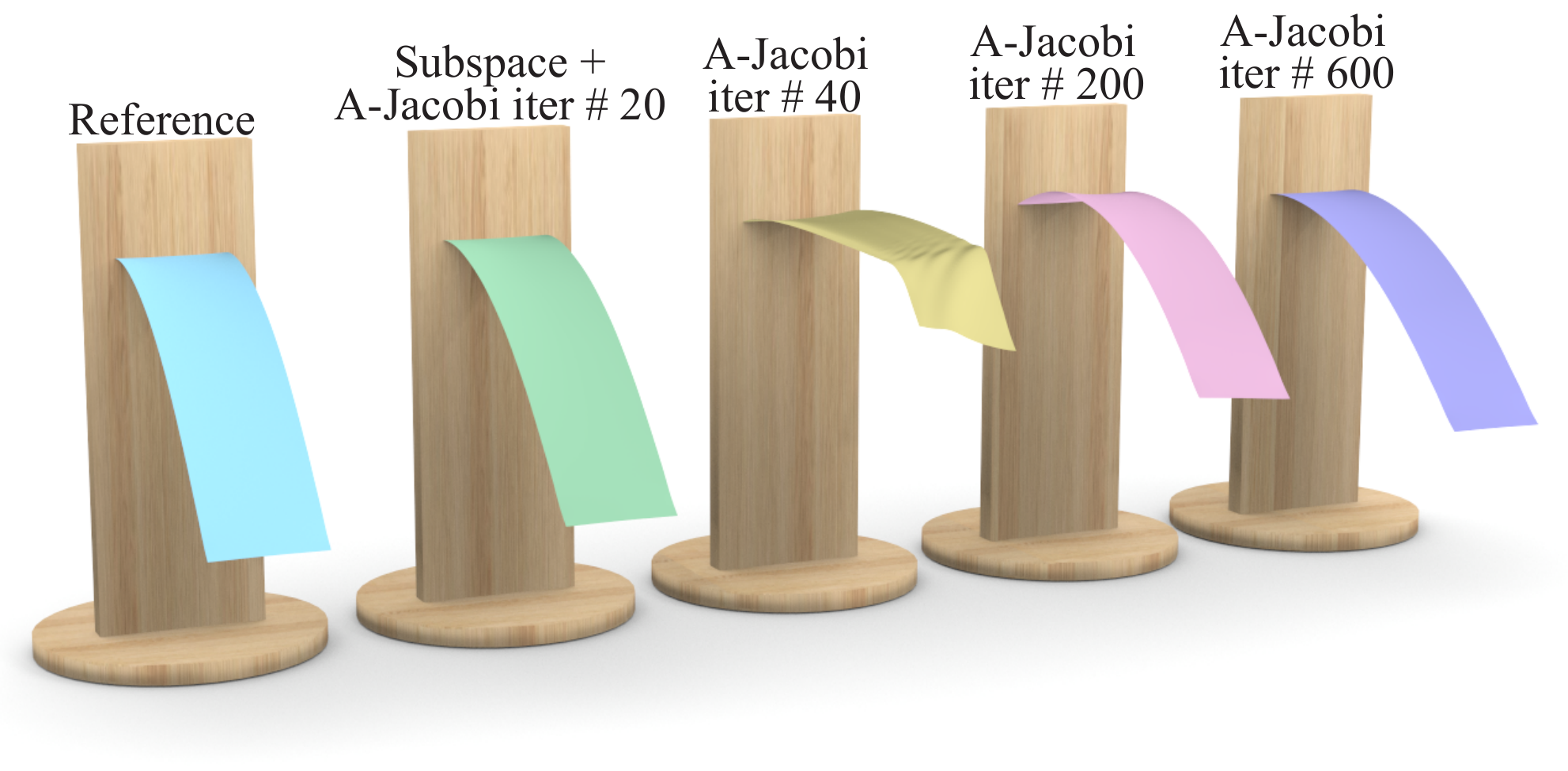}
  \caption{\textbf{Bending strips w. and w/o subspace.}~~We simulate a collision-free scene where a cloth strip bends under gravity. The model consists of $30$K DOFs. The resulting deformation is of low frequency, which is challenging for Jacobi-like methods. A subspace solve effectively resolves this issue: only $20$ A-Jacobi iterations are needed to fully converge the simulation, which otherwise takes over one thousand iterations. Because the bending stiffness is quite strong in this example, we need to assign a big SOR-like weight ($\omega = 0.9$) to dampen each A-Jacobi iteration. Our method runs over $300$ FPS for this example, while PD-IPC~\cite{lan2022penetration} is less than 0.5 FPS due to the large number of A-Jacobi iterations.}
  \label{fig:beam}
\end{figure}

\begin{figure}
  \includegraphics[width=1.0\linewidth]{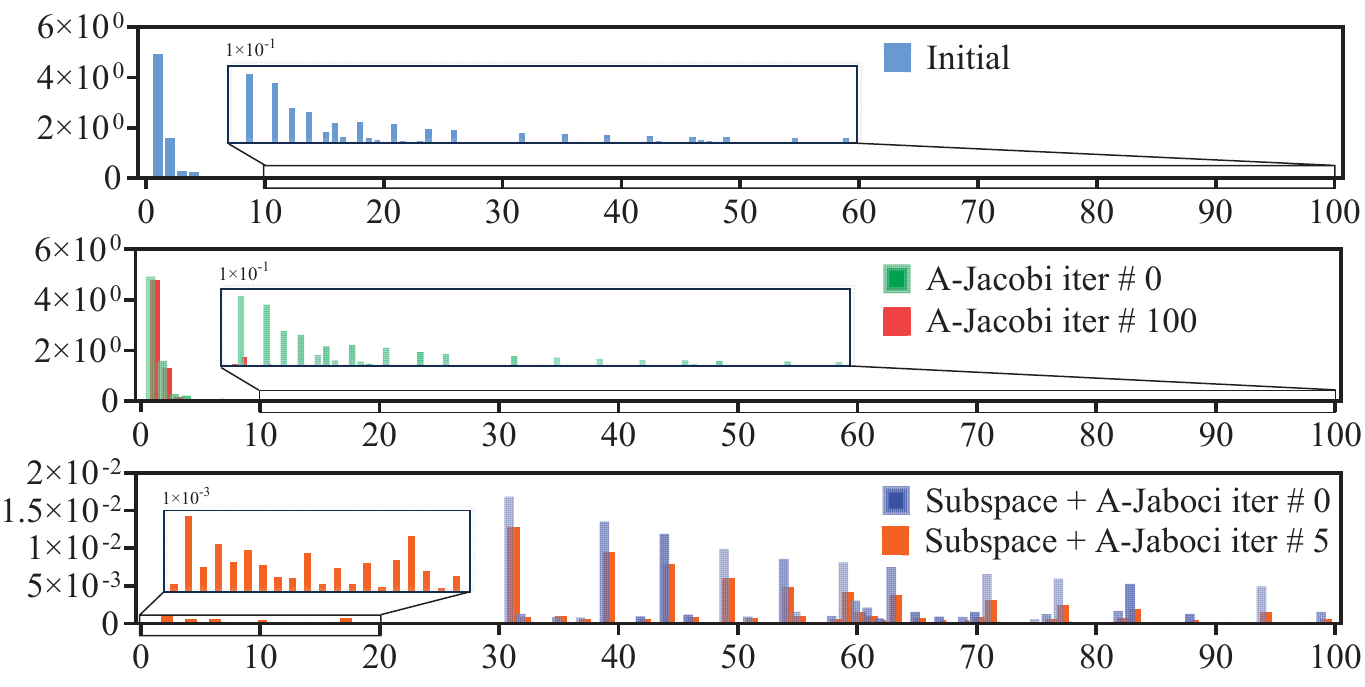}
  \caption{\textbf{Spectral distribution of residual errors (w/o collision).}~~We plot the distribution of residual error over the first 100 modal bases of the strip test (Fig.~\ref{fig:beam}). As shown at the top, the dominant deformation is low-frequency, which is efficiently solved within the subspace (bottom). On the other hand, A-Jacobi iterations are not effective in dealing with low-frequency residuals. 100 A-Jacobi iterations barely lower the low-frequency errors, while the high-frequency strains are well relaxed (middle).}
  \label{fig:error_projection}
\end{figure}

NDB enables a more affordable collision processing, but it does not alter the fact that the optimization of Eq.~\eqref{eq:var_barrier} is highly nonlinear.  Solving the system within a limited time budget remains a computational challenge for time-sensitive applications. Under the framework of PD, the bottleneck is the linear solve at the global stage i.e., Eq.~\eqref{eq:global}. Commonly used strategies resort to iterative GPU solvers such as Jacobi~\cite{wang2015chebyshev} or Gauss-Seidel~\cite{fratarcangeli2016vivace} to solve Eq.~\eqref{eq:global} inexactly. Unfortunately, the presence of barriers makes those solvers less beneficial. Specifically, it is known that iterative solvers are effective in smoothing high-frequency errors but become cumbersome when dealing with low-frequency residuals. In contrast, low-frequency cloth deformation can be efficiently handled using subspace methods. A concrete example is shown in Figs.~\ref{fig:beam} and~\ref{fig:error_projection}. To this end, we tackle Eq.~\eqref{eq:global} using the reduced direct solver and fullspace iterative solver, aiming to reap the goods from both sides. 

Our subspace is for the global stage only. While building a reduced model for the local step is also possible~\cite{brandt2018hyper}, we explicitly avoid doing so to retain local details like wrinkles, folds, and creases. The subspace matrix $\mathsf{U}$ is made of eigenvectors of l.h.s. of Eq.~\eqref{eq:global}, corresponding to $r$ smallest eigenvalues. We first solve Eq.~\eqref{eq:global} in the column space of $\mathsf{U}$ for the reduced displacement $q$:
\begin{equation}\label{eq:subspace_global}
\begin{aligned}
\mathsf{U}^\top \mathsf{H} \left(X + \mathsf{U} q\right) = \mathsf{U}^\top b &\;\Rightarrow\; \left(\mathsf{U}^\top \mathsf{H} \mathsf{U}\right) q  = \mathsf{U}^\top\left(b - \mathsf{H} X\right) \\ &\;\Rightarrow\; 
       \Lambda q  = \mathsf{U}^\top\left(b - \mathsf{H} X\right),
\end{aligned}
\end{equation}
where $\mathsf{H} = \left(\frac{\mathsf{M}}{h^2}+\sum_i w_i\mathsf{S}_i^\top\mathsf{A}_i^\top\mathsf{A}_i\mathsf{S}_i\right)$ and $b = \frac{\mathsf{M}}{h^2} z + \sum_i w_i\mathsf{S}_i^\top\mathsf{A}_i^\top\mathsf{B}_i y_i$ per Eq.~\eqref{eq:global}. $X$ is rest-pose vertex positions. 

As $\mathsf{U}$ is eigenvectors, $\mathsf{U}^\top \mathsf{H} \mathsf{U} = \Lambda$ becomes a diagonal matrix of eigenvalues. If the constraint set stays unchanged during the simulation, solving Eq.~\eqref{eq:subspace_global} is highly efficient on the GPU -- we only need one subspace projection for evaluating $\mathsf{U}^\top(b - \mathsf{H}X)$, and diagonalized subspace solve is negligible. The resulting subspace displacement of $q$ is then converted to its fullspace counterpart as $u = \mathsf{U}q$. At this point, most low-frequency errors have been eliminated by the subspace solve, and $\widetilde{x} = X + \mathsf{U}q$ represents an ideal guess of Eq~\eqref{eq:global} for iterative solvers, which only has high-frequency errors. We then use aggregated-Jacobi (A-Jacobi) as in~\cite{lan2022penetration} to solve for $\Delta x$ to relax the (high-frequency) residual of $\widetilde{x}$:
\begin{equation}\label{eq:high_frequency_residual} 
    \mathsf{H}(\widetilde{x} + \Delta x) = b \;\Rightarrow\; \mathsf{H}\Delta x = b - \mathsf{H}\widetilde{x} \;\Rightarrow\; 
    \mathsf{H}\Delta x = b - \mathsf{H}(X + \mathsf{U}q).
\end{equation}

A representative experiment is reported in Fig.~\ref{fig:beam}, where a cloth strip is attached to the wall and gets bent under gravity. When the low-frequency deformation is computed within the subspace ($r = 30$), it only takes 20 A-Jacobi iterations to converge the simulation to the ground truth i.e., the exact global solve. General A-Jacobi iterations are not effective for such low-frequency deformations -- we observe noticeable visual difference even after 600 iterations (the right-most beam). In this case, the bending stiffness of the cloth is relatively high, and the vanilla Jacobi or A-Jacobi do not even converge. We use a large successive over-relaxation-like (SOR) weight ($\omega = 0.9$) to dampen each A-Jacobi update. The error distribution of this experiment is plotted in Fig.~\ref{fig:error_projection}, which is consistent with our previous analysis.
\begin{wrapfigure}{r}{0.35\linewidth}
    \vspace{-10 pt}
    \includegraphics[width=\linewidth]{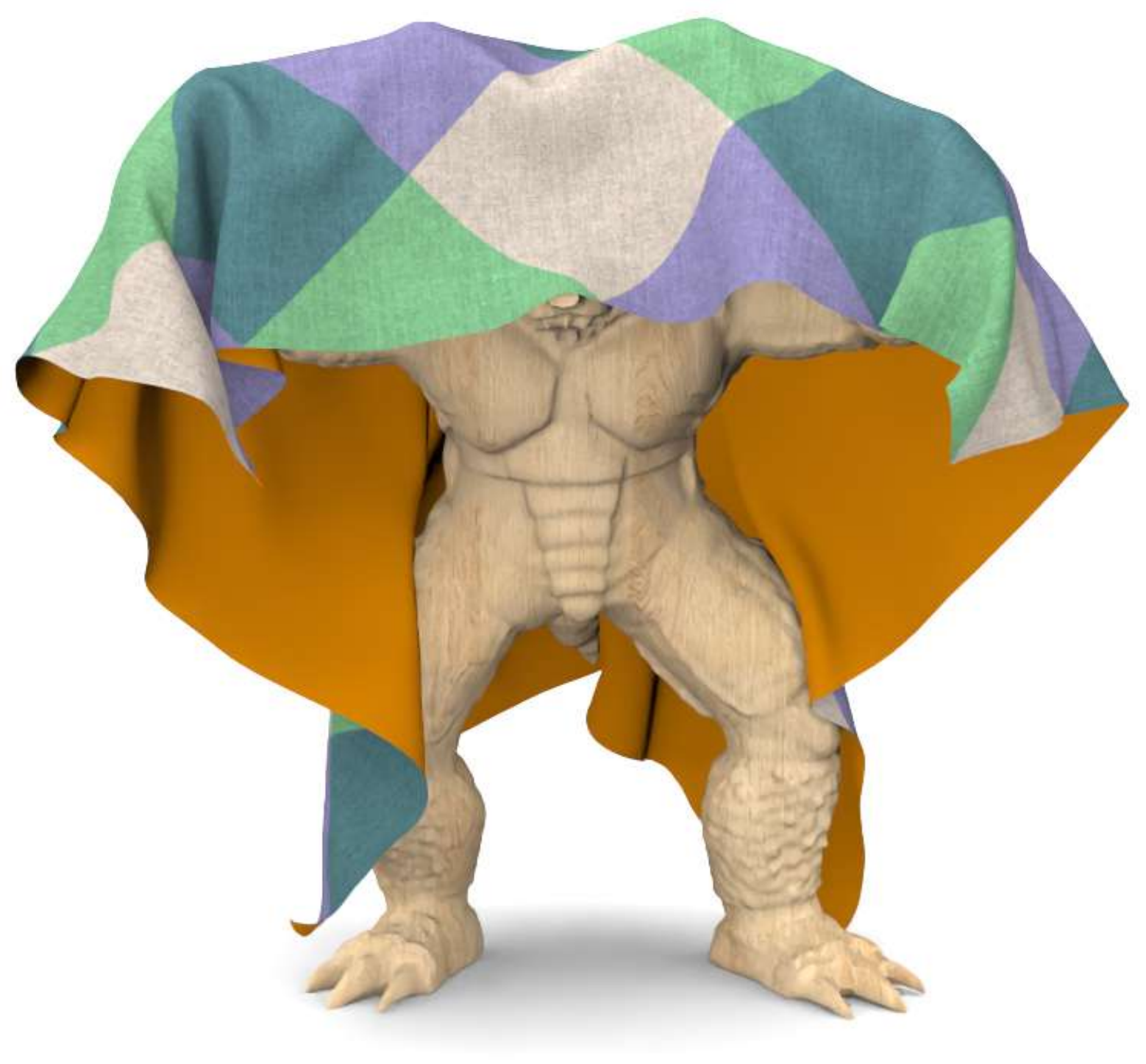}
    
    \caption{\textbf{Subspace reuse.}~~~A piece of tablecloth (66K DOFs) drops on a wooden Armadillo. Over $30\%$ of vertices are involved in collision constraints.}\label{fig:armadillo}
\end{wrapfigure}

Collisions are ubiquitous in cloth simulation, and $\mathsf{H}$ varies during the time integration. Building a new subspace at the simulation runtime does not sound practical for interactive simulations. Our key observation here is: \emph{while the global-stage matrix is altered by different collision constraints, the essential structure of its low-frequency subspace remains unchanged}. It seems counterintuitive at first sight: how can old eigenvectors still be effective while the matrix is modified? This is because low-frequency modes built at the rest shape depict general and global strain distribution over the garment. This is not strongly coupled with high-frequency deformations. For example, the appearance of local wrinkles is less influenced by the overarching movements of the cloth but more by the specific configurations of the corresponding collisions. 

Interestingly, it is the low-frequency deformations that most severely hinder the convergence of iterative solvers. By eliminating or even just reducing these low-frequency errors, we could significantly enhance the convergence of the subsequent Jacobi method.

\begin{figure}
  \includegraphics[width=1.0\linewidth]{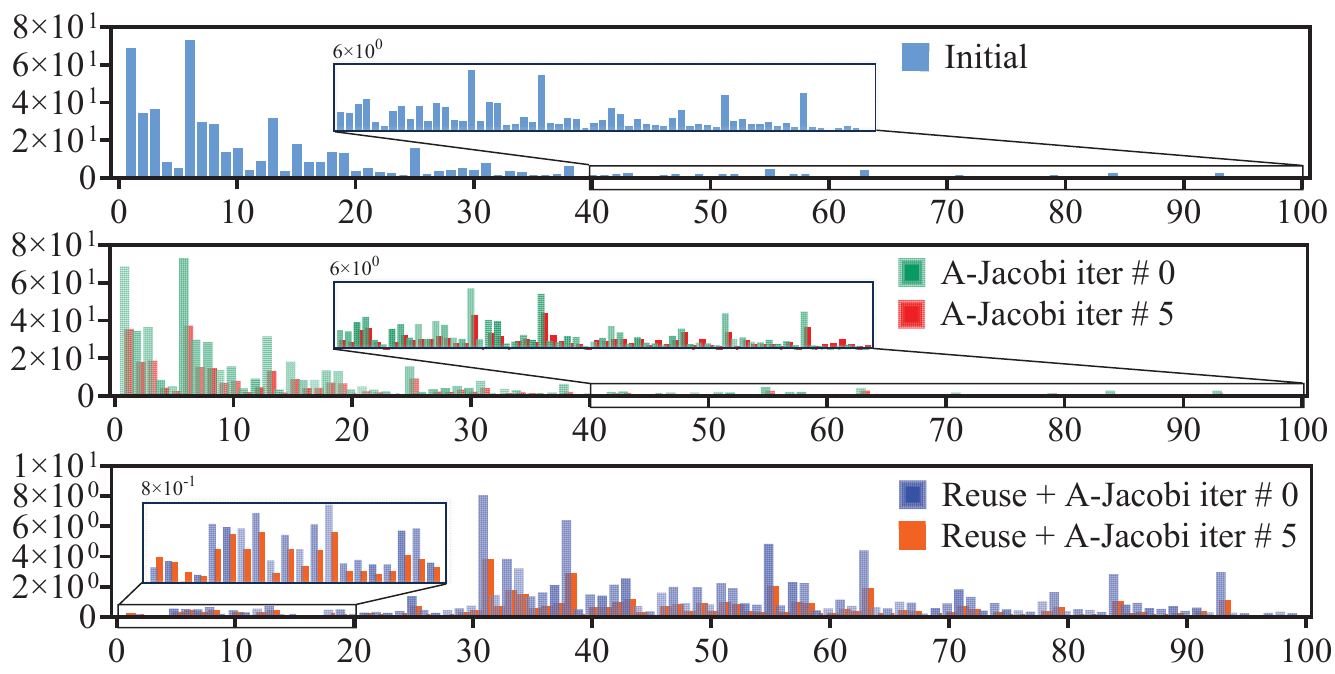}
  \caption{\textbf{Spectral distribution of residual errors (w. collision).}~~We visualize the distribution of residual errors over the first 100 modal bases of the rest-pose global matrix $\mathsf{H}$ when the tablecloth covers the Armadillo (as shown in Fig.~\ref{fig:armadillo}). Subspace reuse does generate some low-frequency errors, but it still helps the convergence of the A-Jacobi significantly.}
  \label{fig:error_projection_subspace_reuse}
\end{figure}

\begin{figure}
  \includegraphics[width=\linewidth]{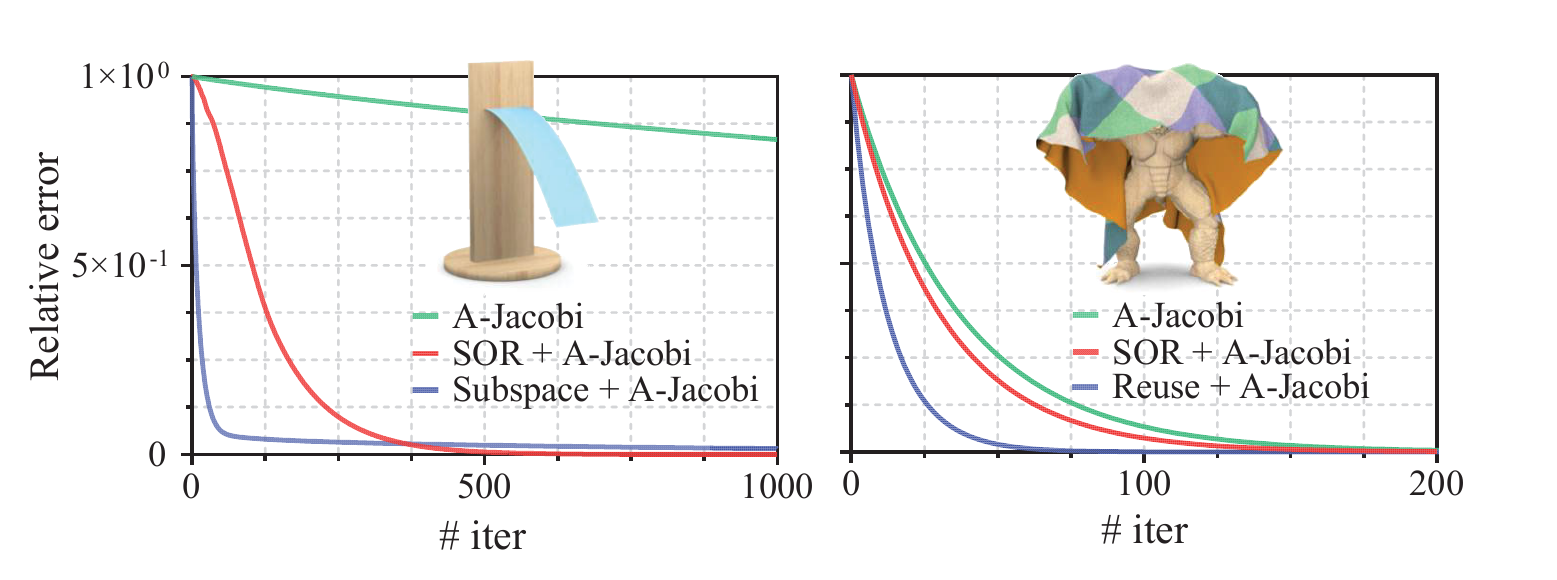}
  \caption{\textbf{Convergence curves w. and w/o subspace (reuse).}~~We plot the convergence curves for experiments in Figs.~\ref{fig:beam} and \ref{fig:armadillo}. We can see subspace solve saves a large faction of A-Jacobi iterations even under intensive collisions. The computation cost of one subspace solve, on the other hand, is similar to performing one or two A-Jacobi iterations on average.}
  \label{fig:convergence_comparision}
\end{figure}

Following this rationale, our subspace reuse strategy pre-computes $\mathsf{U}$ as eigenvectors of the smallest $r$ eigenvalues of $\mathsf{H}$ at the rest pose. During the simulation, the new global matrix becomes $\mathsf{H} + \Delta \mathsf{H}$ i.e., $\Delta \mathsf{H}$ is the modification of $\mathsf{H}$ induced by collisions or self-collisions. Despite the matrix change, we stick with rest-shape subspace and solve $q$ out of the following reduced global-stage system: 
\begin{equation}\label{eq:subspace_collision}
    \mathsf{U}^\top(\mathsf{H}+\Delta \mathsf{H})\mathsf{U} q = \left(\Lambda + \mathsf{U}^\top\Delta \mathsf{H}\mathsf{U}\right) q = \mathsf{U}^\top (b - \mathsf{H}X - \Delta\mathsf{H}X).
\end{equation}
Solving the above system is slightly more expensive than Eq.~\eqref{eq:subspace_global} since $\left(\Lambda + \mathsf{U}^\top\Delta \mathsf{H}\mathsf{U}\right)$ is no longer diagonal. For a subspace of low dimension e.g., $r = 30$, it is still quite efficient using less than $0.1~ms$. Fig.~\ref{fig:armadillo} shows an experiment where a piece of tablecloth covers a wooden Armadillo. The cloth has $66$K DOFs, and over $25$K DOFs are associated with collision constraints. Similar to Fig.~\ref{fig:error_projection}, we report the error distribution for this simulation, where the subspace bases are constructed with the rest-pose matrix $\mathsf{H}$. Clearly, reused subspace solve is less perfect compared with Fig.~\ref{fig:error_projection}. It still effectively handles low-frequency errors, which will need several hundred A-Jacobi iterations otherwise. The convergence curves using subspace and subspace reuse are reported in Fig.~\ref{fig:convergence_comparision}.

\subsection{Pre-computed subspace update}
The most expensive computation is the subspace projection of $\Delta \mathsf{H}$ i.e., evaluating $\mathsf{U}^\top\Delta \mathsf{H}\mathsf{U}$. The complexity is $O(N^2r)$ on the surface. This is prohibitive even on the GPU if we want the simulation to be interactive and high-resolution at the same time.  

We obviate this difficulty exploiting the unique structure of $\mathsf{U}^\top\Delta \mathsf{H}\mathsf{U}$. First, we set $\mathsf{A}_i = \mathsf{B}_i = \mathsf{Id}$ in Eq.~\eqref{eq:local} as an identity matrix ($\mathsf{Id}$) for local projection of collision constraints whose target (collision-free) positions are computed as in~\cite{lan2022penetration}. With NDB, the target positions can be achieved effectively within just a few LG iterations. We note that $\Delta \mathsf{H}$ becomes a diagonal matrix under this treatment.  

Suppose that there are $k$ colliding vertices indexed as $c_1$, $c_2$, $\cdots$, $c_k$, and the weights of the corresponding collision constraints are $w_1$, $w_2$, $\cdots$, $w_k$ respectively. Let us denote each $r$-dimension row vector of $\mathsf{U}$ as $U_i^\top$ such that $\mathsf{U} = \left[U_1, U_2,...\right]^\top$. It can be verified that:
\begin{equation}\label{eq:uu}
    \mathsf{U}^\top \Delta\mathsf{H} \mathbf{U} = \sum_{j = 1}^k w_j U_{c_j} \otimes U_{c_j}\in\mathbb{R}^{r \times r}.
\end{equation}
Therefore the change of the subspace matrix caused by those $k$ colliding vertices is the weighted summation over $k$ rank-one matrices. All of these $U_{c_j} \otimes U_{c_j}$ can be pre-computed, and every $U_{c_j} \otimes U_{c_j}$ is a $r \times r$ symmetric matrix. In practice, we only save its upper triangle at each vertex, and the corresponding memory footprint is lightweight. The computation of the summation, on the other hand, is unfolded at each of $r (r - 1) / 2$ matrix elements using \texttt{CUDA}. The pre-computed subspace update using Eq.~\eqref{eq:uu} is $50$ to $150$ times faster than directly evaluating $\mathsf{U}^\top \Delta \mathsf{H} \mathsf{U}$ using \texttt{cuBlas}. 

After $q$ is computed, we use A-Jacobi to relax the remaining high-frequency residual as in Eq.~\eqref{eq:high_frequency_residual} and move to the next LG iteration until the stopping criterion is satisfied. After an LG iteration, partial CCD ensues, and $w_i$ of the collision constraints are updated per Eq.~\eqref{eq:exp}. Our subspace reuse scheme can efficiently calculate the updated (reduced) global matrix and allows the simulation to enjoy the advantages of both subspace solvers and iterative solvers with small computational costs. Such combined efficiency and convergence are unseen in previous GPU algorithms (e.g., see Fig.~\ref{fig:convergence_comparision}). In practice, we use two subspaces to handle simulation without and with collisions (as detailed in Sec.~\ref{subsec:implementation}).  It is noteworthy that this method is also highly effective for volumetric deformable models, where low-frequency motions are particularly challenging for iterative solvers.

\section{Residual Forwarding}\label{sec:forward}
In interactive applications, simulation modules normally have prescribed time budgets regardless if the solver reaches the convergence. This hard constraint forces the simulator to enter the final CCD-based linear search filtering and truncate the position update $\Delta x$ by the global TOI: $\Delta x \leftarrow \alpha t_{TOI} \cdot \Delta x $. If the TOI is a smaller quantity, severe damping or locking artifacts could be produced as a considerable portion of the system energy dissipates by the early termination. 

To \emph{partially} alleviate this issue, we propose a post-step treatment namely residual forwarding, or RF in short. The idea of RF is to estimate the remaining residual generated by small-step line search filtering and/or non-convergent LG iterations. Recall that each time step, the simulation seeks a minimizer $x^\star$ of Eq.~\eqref{eq:var_barrier}, which ideally should possess a vanished gradient $\nabla E(x^\star) = 0$. By the end of a time step, if LG iterations fail to fully converge, the resulting $x^*$ is different from $x^\star$ such that $x^\star = x^* + \delta x$. The gradient of the variational energy $\nabla E(x^*)= -f_r \neq 0$ represents unbalanced residual forces in the system. RF seeks the virtual force $\delta f$ at the next time step to mitigate the damping artifacts induced by $\delta x$. Therefore, Eq.~\eqref{eq:z} becomes:
\begin{equation}\label{eq:z_rf}
    z = x^* + h \dot{x}^* + h^2 \mathsf{M}^{-1} (f_{{ext}} + \delta f).
\end{equation}
On the other hand, if the actual minimizer $x^\star = x^* + \delta x$ were used, the ground truth $z$ should be:
\begin{equation}
z = x^\star + h \dot{x}^\star + h^2 \mathsf{M}^{-1} f_{{ext}} = x^* +  h \dot{x}^* + 2\delta x + h^2 \mathsf{M}^{-1} f_{{ext}}.
\end{equation}
Note $\dot{x}^\star = \dot{x}^* + \frac{\delta x}{h}$ also depends on the previous position under implicit Euler.
By adding $\delta f$, RF offsets the derivation of $2\delta x$ with the compensation of $h^2 \mathsf{M}^{-1} \delta f$.

To compute the optimal $\delta f$, we Taylor expand $\nabla E$ around $x^\star$ as:
\begin{equation}
\nabla E (x^\star) =  \nabla E(x^*) + \nabla^2 E(x^*) \cdot \delta x + \epsilon(\|\delta x\|^2),
\end{equation}
where $\epsilon$ is a quadratic error term. Because $\nabla E (x^\star) = 0$, we obtain:
\begin{equation}\label{eq:del_x}
    \delta x  = \left(\nabla^2 E(x^*)\right)^{-1}\left( f_r - \epsilon(\|\delta x\|^2)  \right).
\end{equation}
Assuming $\epsilon(\|\delta x\|^2)$ is sufficiently small, the most effective RF should minimize:
\begin{equation}\label{eq:rf}
    \arg \min_{\delta f} \left\| \frac{h^2}{2}\mathsf{M}^{-1} \delta f - \left(\nabla^2 E(x^*)\right)^{-1} f_r \right\|.
\end{equation}
Taking a closer look, it is noted that computing $\left(\nabla^2 E(x^*)\right)^{-1} f_r$ is equivalent to taking one more Newton solve by the end of the previous time step, which can be efficiently approximated with subspace reuse. In RF, \emph{we do not need to perform CCD-based line search filtering}, and the weights for all the collision constraints is $k$ i.e., $a_i$ for $i$-th collision constraint is set zero in Eq.~\eqref{eq:exp}.

\paragraph{Discussion}
RF is a \emph{heuristic treatment} when the current time step must end for other time-critical tasks. If $\epsilon(\|\delta x\|^2)$ is big, RF becomes erroneous and generates artifacts. Honestly, there is nothing much we can do if the available time budget is aggressively restrained. Conceptually, RF moves some computation e.g., solve for $\left(\nabla^2 E(x^*)\right)^{-1} f_r$ to the next time step. The advantage of RF is skipping the line search filtering since $z$ could embody an overlapping and penetrating configuration. The collision constraint set is then updated in the follow-up LG iteration in the future time step. RF, on the other hand, allows non-colliding vertices to move under cloth momentum and elasticity even with a small $t_{TOI}$ (i.e., from the previous time step). As a result, damping/locking artifacts are ameliorated.

\begin{figure*}
  \centering
  \includegraphics[width=\linewidth]{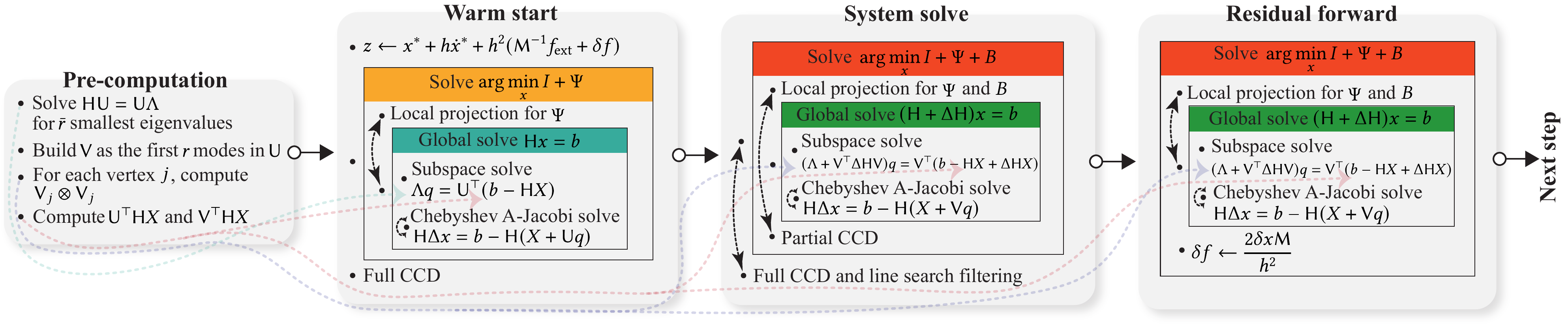}
  \caption{\textbf{Algorithm overview.}~~Our method leverages a subspace reuse technique to improve the convergence of GPU-based iterative solvers. A large portion of costly CCDs is replaced with a non-distance non-distance barrier formulation i.e., NDB. The pipeline also features a residual forwarding mechanism, which generates plausible animation even with small exiting TOI. We use a bigger subspace of $\bar{r}$ dimension to obtain a good warm start at the beginning of each time step. A more compact $r$-dimension subspace is used for handling NDB-in-the-loop optimization. Each time step consists of several blocks of LG iterations, which are visualized with curved double arrows.}
  \label{fig:flowchart}
\end{figure*}

\setlength{\textfloatsep}{10 pt}
\begin{algorithm}
\SetNlSty{small}{}{:}
$z \leftarrow x^* + h\dot{x}^* + h^2(\mathsf{M}^{-1} f_{ext} + \delta f)$ \tcp{$\delta f$ is from RF}
$x \leftarrow z$\\
$\|\Delta x\|\leftarrow \infty$\\
\While {$\|\Delta x\| < \epsilon_{initial}$}
{
    $\Delta x \leftarrow$ \textsf{ModalLG}$(I + \Psi)$  \tcp{in $\text{span}(\mathsf{U})$}
}
$x^-\leftarrow x^* $\\
$x^+\leftarrow x + \Delta x$\\
$\langle B, t_{TOI}\rangle \leftarrow$ \textsf{FullCCD}$(x^-, x^+)$\tcp{$B$ is the latest barrier}
$x \leftarrow x + t_{TOI} \cdot \Delta x$\\
$x^-\leftarrow x$\\
\While {outer loop convergence check fails}
{
    \While{inner loop convergence check fails}
    {
        $\Delta x \leftarrow$ \textsf{ModalLGReuse}$\left(I + \Psi + B \right)$  \tcp{in $\text{span}(\mathsf{V})$}
        $x^+\leftarrow x + \Delta x$\\
        $B_i \leftarrow$\textsf{PartialCCD}$(x^-, x^+)$
    }
    $\langle B, t_{TOI}\rangle \leftarrow$ \textsf{FullCCD}$(x^-, x^+)$ \tcp{update $B$ at each outer loop}
}
$x \leftarrow x + t_{TOI} \cdot \Delta x$\tcp{exiting line search filtering}
\If {$t_{TOI} < \epsilon_{TOI}$}
{
    $B_i = \kappa$ \tcp{quadratic approximation of $B$}
    \While{$\|\Delta x\| < \epsilon_{\text{inner}}$}
    {
        $\delta x \leftarrow$ \textsf{ModalLGReuse}$\left(I + \Psi + B\right)$ \\ 
    }
    $\displaystyle \delta f \leftarrow \frac{2\delta x \mathsf{M}}{h^2}$
}
\caption{Our simulation pipeline.}\label{alg:m2}
\end{algorithm}

\section{Simulation Pipeline}\label{sec:pipeline}
We now have all the pieces to assemble our simulator. Fig.~\ref{fig:flowchart} visualizes major steps along our pipeline, and the pseudocode is also outlined in Alg.~\ref{alg:m2}. The pre-computation stage performs the eigendecomposition of the rest-shape global matrix $\mathsf{H}$, which ignores the collision constraints.
We use two subspaces to handle global solves without and with collisions. Specifically, a subspace of higher dimension is first constructed out of $\bar{r}$ eigenvectors. Because the rest-shape modal global matrix $\Lambda = \mathsf{U}^\top \mathsf{H} \mathsf{U}$ is diagonal, and the subspace projection of r.h.s. vector is efficient on GPU, increasing the dimensionality of the subspace for collision-free global solve is worthy and effective. The first $r < \bar{r}$ eigenvectors of $\mathsf{U}$ form a smaller set of bases $\mathsf{V}$ of a more compact subspace, and we set $r = 30$ in our implementation. $\mathsf{V}$ is for subspace reuse when collision constraints are taken into account. For each vertex $j$, we pre-compute the corresponding ${V}_j\otimes {V}_j$ as in Eq.~\eqref{eq:uu} for fast computation of the updated subspace matrix. Lastly, $\mathsf{U}^\top\mathsf{H}X$ and $\mathsf{V}^\top\mathsf{H}X$ are also pre-computed for faster assembly of the r.h.s. of the global solve (i.e., Eqs.~\eqref{eq:subspace_global} and \eqref{eq:subspace_collision}).

\begin{figure*}
  \centering
  \includegraphics[width=\linewidth]{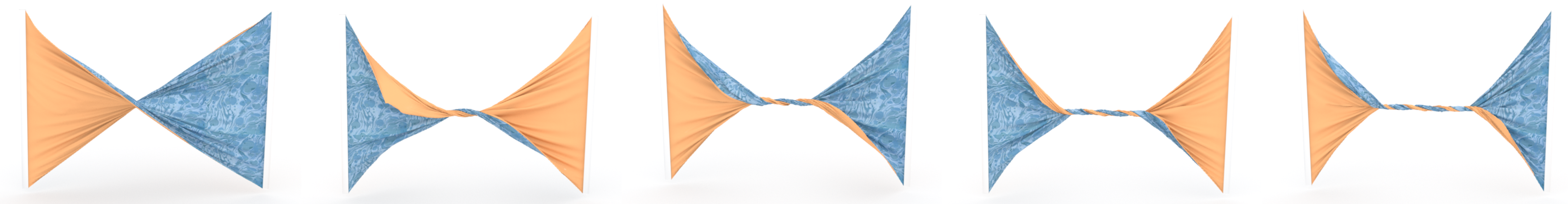}
  \caption{\textbf{Cloth twisting NDB.}~~We rotate the table cloth at both ends for $2,880$ degrees. The simulation includes over $120$K DOFs. Our method completes this test at 12 FPS.  Compared with the distance-based barrier weighted PD method i.e.,~\cite{lan2022penetration}. NDB saves about $30\%$ LG iterations (as plotted in Fig.~\ref{fig:exp_dbb}).} 
  \label{fig:twist}
\end{figure*}

Keeping $\mathsf{V}$ condensed is helpful for both efficiency and efficacy of the simulation. From the efficiency point of view, we know $\Lambda + \mathsf{V}^\top\Delta \mathsf{H}\mathsf{V}$ is not diagonal. Solving a dense linear system is only feasible when the system is of low dimension. From the efficacy point of view, the subspace reuse is effective in the low-frequency realm. Barrier constraints induced by collisions and contacts can drastically change the higher-frequency landscape. As a result, expanding $\mathsf{V}$ to the high-frequency spectrum is not beneficial. Since we will need to save ${V}_j \otimes {V}_j$ for each vertex, a more compact ${V}_j$ is also more memory-friendly. 

Each time step begins with a warm-start computation. Typically, $z$ (i.e., in Eq.~\eqref{eq:z_rf}) is used as the initial guess of $x$ at the current time step and for setting up the initial constraints $B^{NDB}$. As the collision is ignored for the warm start, $\bar{r}$-dimension subspace provides a better initialization by solving $\arg\min_x I + \Psi$ (line 5 in Alg.~\ref{alg:m2}). The decoupled computations in the modal space of span($\mathsf{U}$) make this procedure highly efficient. For instance, the warm start only takes three to five LG iterations and less than five milliseconds for a $300$K-DOF simulation. The resulting $x$ is forwarded to the full CCD procedure. The constraint list is then built, and $B(x)$ is initialized (line 9). 

Afterwards, the system solves for the optimization of $\arg\min_x I + \Psi + B$. Similar to existing barrier-in-the-loop PD algorithms~\cite{lan2022penetration}, we employ a two-level iteration scheme. The outer loop consists of multiple LG iterations, and the outer convergence check is based on the norm of the change of $x$ between two consecutive outer loops. If we uniformly scale the cloth meshes to a unit size, $\| \Delta x \| \leq 1E-3$ is a good choice for convergence check. For simulations involving fast-moving objects, setting $\| \Delta x \| \leq 5E-4$ may be needed. Each inner loop begins with a standard LG iteration. The local projections are performed for both elasticity constraints ($\Psi$) and collision/barrier constraints ($B$). At the global solve, we re-use the subspace $\mathsf{V}$ to smooth high-frequency errors and pass the residual to A-Jacobi iterations. The convergence of the inner loop is also based on $\|\Delta x \|$ from the previous inner loop ($\|\Delta x \| \leq 5E-2$). The partial CCD routine is then invoked (line 16) to adjust the weights of collision/contact constraints. As discussed in Sec.~\ref{sec:exp}, this computation boils down to computing the inner products of 3-vectors. 

By the end of each time step, a full CCD and an exiting line search filtering are performed. They offer the algorithmic guarantee that the simulation is free of inter-penetration. If the exiting  $t_{TOI}$ is too small suggesting possible locking and overdamping due to insufficient outer loops, we trigger the residual forward, which estimates an optimal correction force $\delta f$ for the next step.

\section{Experimental Results}\label{sec:result}
We implemented the proposed simulation framework on a desktop computer with an \texttt{intel} \texttt{i7-12700} CPU and an \texttt{nVidia} \texttt{3090 RTX} GPU. We used \texttt{Spectra} library for computing the eigendecomposition of the global stage matrix $\mathsf{H}$. It should be noted that our method is also friendly with other parallel computing platforms -- one can easily parallelize local projections using multi-threading, and multicore CPUs are well suited for the Gauss-Seidel method. Nevertheless, we only report the performance on the GPU. The reader can find animated results in the accompanying video demo. 

\subsection{Implementation details}\label{subsec:implementation}
Most parts of our framework are matrix-free, except for the subspace solve step of $\mathsf{V}^\top (\mathsf{H} + \Delta\mathsf{H}) \mathsf{V}$. For solving a linear system of $\mathsf{A} x = b$, the common practice is to pre-factorize $\mathsf{A}$ and compute $x$ via forward and backward substitutions. In our implementation however, we directly obtain $\mathsf{X} \in\mathbb{R}^{r \times r}$ via:
\begin{equation}\label{eq:inverse}
\mathsf{A}\mathsf{X} = \frac{1}{\beta} \mathsf{Id},  
\end{equation}
to get an approximation of $\mathsf{X} \approx \mathsf{A}^{-1}/ \beta$. The reader should not confuse $\mathsf{A}$ in Eq.~\eqref{eq:inverse} with $\mathsf{A}_i$ used in local projection (e.g., Eq.~\eqref{eq:local}). Here, $\mathsf{A}$ refers to the reduced global-stage matrix $\mathsf{V}^\top(\mathsf{H} + \Delta\mathsf{H}) \mathsf{V}$ with subspace reuse. $\beta$ is a scaling factor estimated as $\beta = \sum |b_i|/r$ i.e., the average of absolute values of elements in $b$. Doing so could mitigate numerical drift induced by small or large values in $b$. This is because calculating $\mathsf{X}$ via $\mathsf{AX}=\mathsf{Id}$ implicitly assumes the r.h.s. of the system is around one. 

With $\mathsf{X}$ computed, the system solve becomes a matrix-vector multiplication of $\mathsf{X}b$, which can be parallelized on the GPU. The standard forward/backward substitutions are sequential (even on the GPU). When many LG iterations are needed, computing $\mathsf{X}b$ is more efficient than using the factorized matrix. In our implementation, we simply send $\mathsf{A}$ back to the CPU, compute $\mathsf{X}$, and return it to the GPU. Since the reused subspace has a very low dimensionality, it ensures that the associated computations and CPU-GPU communications are fast and require minimal resources. The whole procedure takes less than $0.1$ ms, which is $30\%-35\%$ faster than factorizing $\mathsf{A}$ on the GPU. Nevertheless, the system solve is \emph{not} the bottleneck. In our implementation, we use rank-2 A-Jacobi method, which computes two regular Jacobi iterations with one step but using the same computation time~\cite{lan2022penetration}. Thanks to subspace reuse, weighted SOR is never needed even for stiff simulations (e.g., Fig.~\ref{fig:beam}). The base of exponential NDB (i.e., Eq.~\eqref{eq:exp}) is set as $K =2 $ in our experiments. Setting $K$ to $3$ seems to produce a similar result. However, an over-aggressive $K$ could negatively impact the convergence. In this case, this exponential NDB behaviors like a geometric projection.

To fully exploit the capacity of modern GPUs, the broad-phase collision culling leverages a patch-based BVH. Specifically, we build an incomplete BVH whose leaf houses a small patch of the cloth mesh. A patch consists of several inter-connected triangles, normally five to eight. After the initial AABB-based culling at each BVH level, from top to bottom, we exhaustively test triangle pairs between two nearby patches as well as pairs within a patch. At the narrow phase stage, if a full CCD is needed we solve $t_{TOI}$ for each primitive pair using the polynomial solver proposed in~\cite{Yuksel2022}. For partial CCD, we compute inner products of Eq.~\eqref{eq:query} at pre-selected sample points plus the projection points at $t = 0$. We periodically check if a denser sampling is needed given the current system velocity (since the time step size is assumed fixed). Partial CCD is more efficient for cloth models of higher resolutions. We noted that, as long as the total number of DOFs exceeds $50$K, very few sample points (e.g., three) work well for partial CCD in most simulation scenarios.

\begin{figure}
 \includegraphics[width=\linewidth]{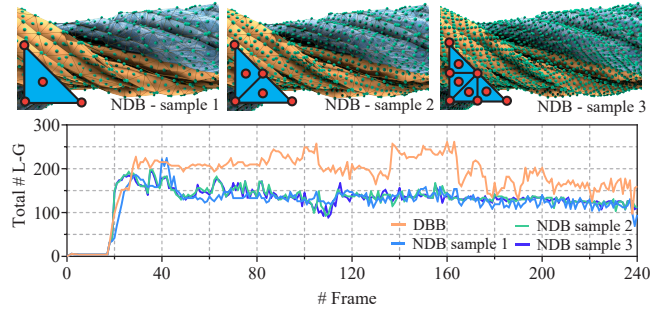}
  \caption{\textbf{NDB vs DBB.} 
  We plot the total number of LG iterations at each time step when twisting the cloth (as in Fig.~\ref{fig:twist}) using DBB and exponential NDB. It can be seen that our NDB formulation uses $25\%$ to $30\%$ fewer iterations on average than DBB when allows the constraints to be adaptively re-weighted in time. Partial CCD is not sensitive to the sample density. As shown in the figure, increasing or decreasing partial CCD samples do not vastly alter the convergence behavior of the simulation. Different sample patterns are also visualized in the figure.}
  \label{fig:exp_dbb}
\end{figure}

\subsection{NDB \& DBB}
The first test we would like to show is a comparison between the distance-based barrier (DBB) and exponential non-distance barrier i.e.,NDB. The major difference lies in the fact that NDB allows the weight of the collision constraint to be timely adjusted during LG iterations based on inexpensive partial CCD. The snapshots of the resulting simulation are reported in Fig.~\ref{fig:twist}. We also compare the total number of LG iterations using NDB and DBB for this twisting test, and the plots are shown in Fig.~\ref{fig:exp_dbb}. It can be seen that this adaptive weighting strategy helps reduce iterations, as each collision constraint is more likely to find an appropriate weight during iterations. NDB strategyis not sensitive to the sample density ($\rho$). To this end, Fig.~\ref{fig:exp_dbb} also plots iteration counts for NDB with different sampling densities. We can see that NDB works well even using one sample point at the center of $\Omega_\lambda$. 
\begin{figure}
  \centering
  \includegraphics[width=\linewidth]{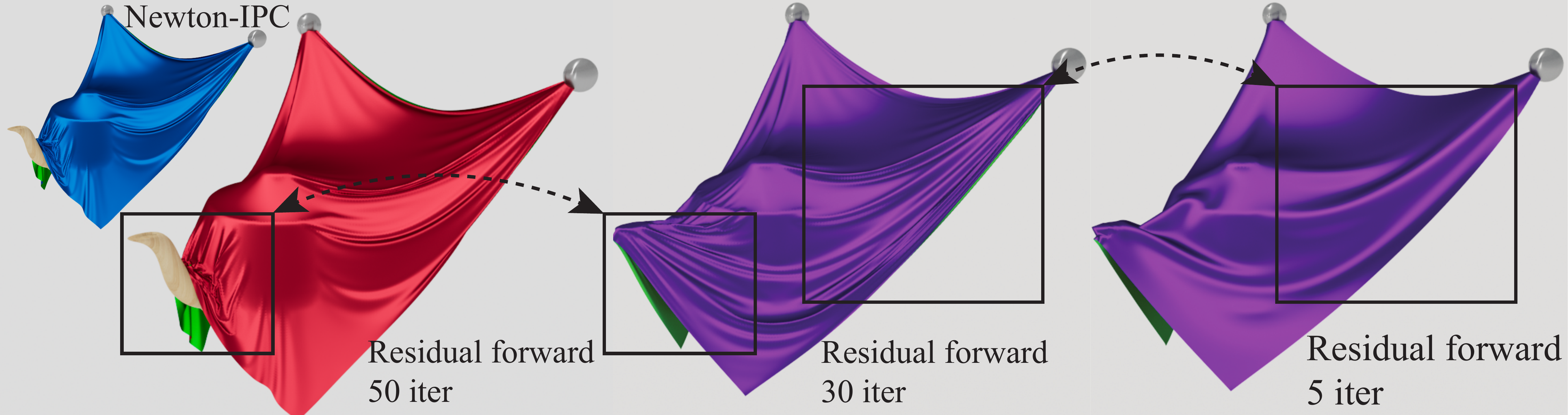}
  \caption{\textbf{Residual forwarding.}~~We show an experiment where a piece of tablecloth falls on the teapot. Residual forwarding estimates a virtual force to restore the dynamics at non-colliding vertices in the next step. It produces reasonably good results when the residual errors are moderate. RF fails to handle all the errors with a highly constrained iteration cap. As we lower the per-step iteration count, we observe more artifacts even using RF.}
  \label{fig:rf}
\end{figure}
\subsection{Residual forwarding}
Our method employs RF to mitigate damping and locking artifacts caused by limited time budgets. As discussed in Sec.~\ref{sec:forward}, RF can enhance animation quality to a certain extent by ``inheriting'' unresolved residual errors from one time step to the next. The effectiveness of RF is primarily up to the magnitude of these errors. We observe that small and localized errors, resulting from early termination or line search filtering under a small TOI, are generally well-managed by RF. Another factor is the capability of minimizing residuals at the current time step. If the solver does not fully converge at the current step, the carried-over residuals may exacerbate simulation inaccuracies. RF is particularly designed for applications facing strict time constraints, such as gaming, where high-speed collisions might cause spiky FPS drops. 

To better illustrate the effectiveness of RF, we simulate a scene where the cloth drops on a teapot (Fig.~\ref{fig:rf}). There are $120$K DOFs on the cloth. After the cloth comes in contact with the spout of the teapot, the inertia effect further moves the cloth towards the right, and the cloth eventually settles at the recess between the spout and the body of the teapot. When we have the time budget to complete 50 iterations per time step (which \emph{do not} fully converge the solver), RF produces a high-quality result nearly identical to the ground truth -- the one generated using a fully converged Newton IPC solver. Our method runs at $20$ FPS, which is $500\times$ faster than~\cite{li2021codimensional}. If we half the iteration number and exit the current time step with the line search filtering, the simulator is unable to deal with all the residuals, and we can see dampened cloth movement after it touches the teapot. Because of the early termination, the cloth ``sticks'' to the tip of the spout as the rest part of the cloth is locked (see highlighted area in the figure). Lastly, we further lower the iteration cap to only five. We observe more severe damping artifacts. Because of the subspace resue, our method still handles low-frequency residuals well, but most high-frequency information is lost due to an insufficient number of A-Jacobi iterations. The entire cloth exhibits rubber-like dynamics, and the collision is highly inelastic. This comparison is available in the supplementary video. 



\begin{figure}
  \centering
  \includegraphics[width=\linewidth]{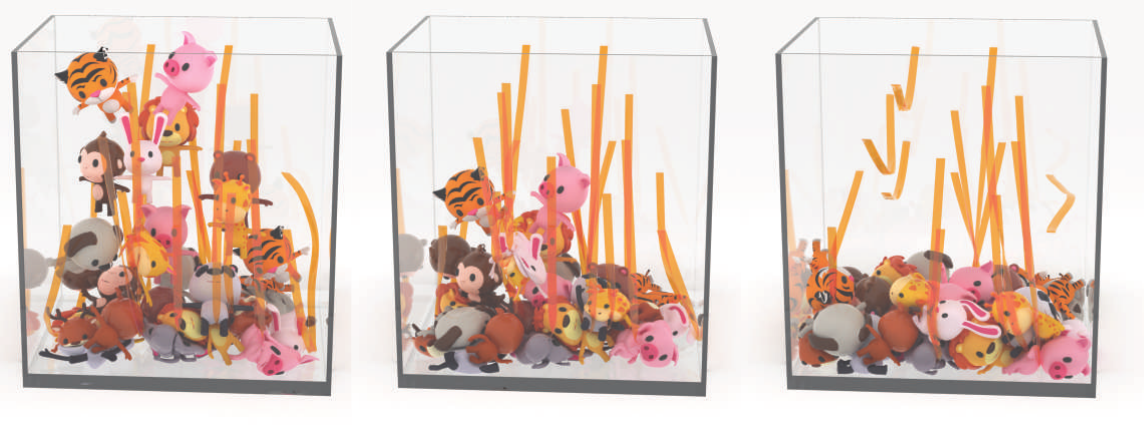}
  \caption{\textbf{``Animal crossing''.}~~The subspace reuse technique allows our solver to handle stiff materials and deformable bodies with ease. This example includes $33$ animal toys, 10 elastic ribbons, $678$K DOFs, and one million elements. Both our method and PD-IPC~\cite{lan2022penetration} produce penetration-free animations, while our method is $12\times$ faster than PD-IPC on \texttt{RTX 3090}.}
  \label{fig:animal}
\end{figure}

\subsection{Comparison with existing methods}
PD is a popular framework, based on which many excellent cloth and deformable simulation algorithms have been developed. In this subsection, we compare our method with several representative PD-based cloth simulation methods, including PD-IPC~\cite{lan2022penetration}, PD-BFGS~\cite{li2023subspace}, and PD-Coulomb~\cite{ly2020projective}. 

\paragraph{Our method vs. PD-IPC}
PD-IPC~\cite{lan2022penetration} is a full GPU simulator. It also integrates PD with IPC for deformable and cloth simulation. Unlike our method, PD-IPC uses DBB, and a full CCD must be invoked every time the constraint set is to be updated. PD-IPC solves the global step system only using A-Jacobi iterations on the GPU. For stiff simulations, one must adopt a close-to-one SOR weight to ensure A-Jacobi iterations do not diverge. This is not an issue for us since the subspace solve removes dominant low-frequency errors. As a result, our method outperforms PD-IPC by a significant margin in general. Fig.~\ref{fig:animal} reports a deformable body simulation result. In this example, we have 33 deformable animal toys, 10 elastic ribbons, and over one million elements falling into a glass tank. Some animal toys are five times stiffer than others. Such heterogeneous materials/bodies are particularly challenging for PD-IPC as the SOR weight of the A-Jacobi must be set conservatively ($\omega \geq 0.85$), which impairs the convergence. In this example, our method is $12\times$ faster. As NDB does not depend on the distance, updating NDB can be processed using partial CCD. This strategy significantly speeds up the evaluation of barrier weighting and reduces more than 60\% of computation time. 

\begin{figure}
  \centering
  \includegraphics[width=\linewidth]{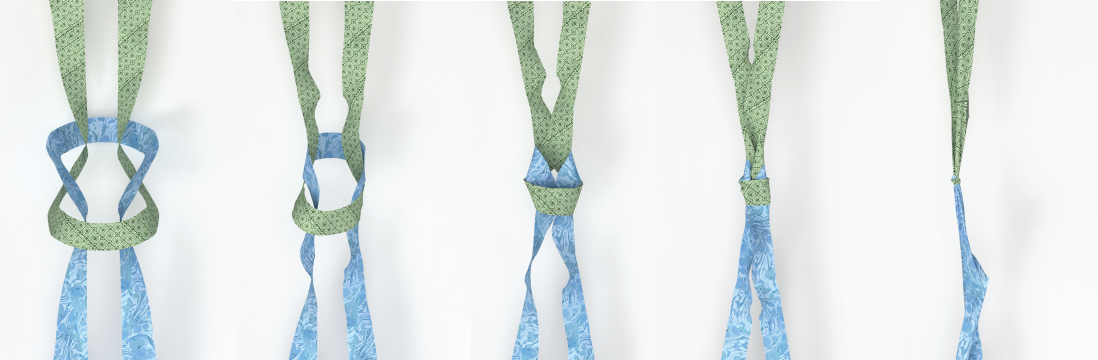}
  \caption{\textbf{Make a knot.}~~Two cloth strips are pulled from opposite directions to form a tight knot. Both our method and PD-BFGS~\cite{li2023subspace} successfully handle this challenging simulation. Nevertheless, our method is $130\times$ faster due to sample-based partial CCD with exponential NDB re-weighting strategy, and a more efficient subspace reuse strategy on global solve. In this experiment, there are $104$K vertices and $203\text{K}$ triangles on the mesh.}
  \label{fig:knot}
\end{figure}

\begin{table*}
\caption{\textbf{Experiment statistics.}~~We report detailed time statistics for experiments mentioned in the paper. \textbf{$\#$ B} gives the total number of objects in the scene. \textbf{$\#$ DOF} is the total number of simulation DOFs. \textbf{$\#$ Ele.} records the total number of elements (i.e., triangles and tetrahedrons). \textbf{$\#$ Con.} and \textbf{$\#$ Col.} are average numbers of elasticity constraints and collision constraints during the simulation. In the column of \textbf{$\#$ Col.}, the first quantity reports the number of collision constraints involved in CCD and the second number is the total number of primitive pairs after the broad phase collision detection.  $\bm{\bar{r} | r}$ represent the subspace size and reused subspace size.  $\bm{\Delta t}$ is the time step size. \textbf{Pre.} is the pre-computation time (measured in seconds). The column \textbf{B|N|P|F} gives the timing information (in milliseconds) used for collision detection. Specifically, \textbf{B} is the broad phase time, and \textbf{N} is the narrow phase time. The narrow phase also includes partial CCD (\textbf{P}) and full CCD (\textbf{F}) procedures. $\rho_m$ gives the mass density of the cloth (and the deformable objects e.g., in Figs.~\ref{fig:animal} and~\ref{fig:cover_ship}). The unit is $\frac{kg}{m^2}$ for the cloth (and $\frac{kg}{m^3}$ for deformable objects). $\kappa$ (in $MPa$) gives the stretching and bending stiffness of the cloth (and Young's modulus of deformable objects). $\|\Delta x\|$ is the convergence condition. \textbf{LG ($\#$)} gives the total time (first row) used and total number of LG iterations (second row) on average for each step. \textbf{Misc.} corresponds to other additional computations. Except for $\textbf{Pre.}$, other timings are measured in milliseconds. \textbf{FPS} is the overall FPS of the simulation.}\label{tab:time}
{\footnotesize \fontfamily{ppl}\selectfont
\begin{center}
\begin{tabular}{c||c|c|c|c|c|c|c|c|c|c|c|c|c|c|c}
\whline{1.15pt}
\textbf{Scene} & \textbf{$\#$ B} & \textbf{$\#$ DOF} & \textbf{$\#$ Ele} & \textbf{$\#$ Con.} & \textbf{$\#$ Col.} & \textbf{$\bm{\bar{r} | r}$} & $\bm{\Delta t}$ & \textbf{Pre.} &\textbf{B|N|P|F} & $\rho_m$ & $\kappa$ & $\|\Delta x\|$ & \textbf{LG ($\#$)}  & \textbf{Misc.} & \textbf{FPS} \\
\whline{0.5pt}
\begin{tabular}{@{}c@{}} \scriptsize{Twisting cloth} \\ \scriptsize{(Fig.~\ref{fig:twist})}  \end{tabular}
& $1$ & $121\text{K}$ & $80\text{K}$ & $161\text{K}$ & \begin{tabular}{@{}c@{}} $110\text{K}$ \\ $8\text{M}$  \end{tabular} & $120|30$ & $\frac{1}{150}$ & $6.3$ & $8|15|2|13$ &$0.3$ & \begin{tabular}{@{}c@{}} $160$ \\ $3\cdot10^{-4}$  \end{tabular} & $1\cdot10^{-3}$  & \begin{tabular}{@{}c@{}} $47.4$ \\ ($157$)  \end{tabular} & $10.2$ & $12.4$ \\

\whline{0.5pt}
\begin{tabular}{@{}c@{}} \scriptsize{``Animal cross.''} \\ \scriptsize{(Fig.~\ref{fig:animal})}  \end{tabular}
& $44$ & $678\text{K}$ & $1.1\text{M}$ & $1.1\text{M}$ & \begin{tabular}{@{}c@{}} $67\text{K}$ \\ $9.2\text{M}$  \end{tabular} & $120|30$ & $\frac{1}{150}$ & $46.8$ & $20|10|0.4|10$  &$0.9$ & $2\cdot 10^3$ &$1\cdot10^{-3}$   & \begin{tabular}{@{}c@{}} $103.2$ \\ ($47$)  \end{tabular} &$9.2$ & $7.0$ \\

\whline{0.5pt}
\begin{tabular}{@{}c@{}} \scriptsize{Make a knot} \\ \scriptsize{(Fig.~\ref{fig:knot})}  \end{tabular}
& $1$ & $310\text{K}$ & $203\text{K}$ & $410\text{K}$ & \begin{tabular}{@{}c@{}} $13\text{K}$ \\ $4\text{M}$  \end{tabular} & $120|30$ & $\frac{1}{150}$ & $13.6$ & $8|7|0.2|7$ &$0.3$ &\begin{tabular}{@{}c@{}} $160$ \\ $3\cdot10^{-4}$  \end{tabular} &$1\cdot10^{-3}$  & \begin{tabular}{@{}c@{}} $63.6$ \\ ($74$)  \end{tabular} & $6.1$ & $11.8$\\

\whline{0.5pt}
\begin{tabular}{@{}c@{}} \scriptsize{Drape one a sphere} \\ \scriptsize{(Fig.~\ref{fig:pd_coulomb})}  \end{tabular}
& $2$ & $120\text{K}$ & $80\text{K}$ & $160\text{K}$ & \begin{tabular}{@{}c@{}} $7\text{K}$ \\ $3\text{M}$  \end{tabular} & $120|30$ & $\frac{1}{200}$ & $6.2$ & $4|4|0.2|4$ &$0.3$ &\begin{tabular}{@{}c@{}} $100$ \\ $2\cdot10^{-4}$  \end{tabular} &$1\cdot10^{-3}$  & \begin{tabular}{@{}c@{}} $30.8$ \\ ($38$)  \end{tabular} & $1.2$ & $25.0$ \\

\whline{0.5pt}
\begin{tabular}{@{}c@{}} \scriptsize{Stack on teapot} \\ \scriptsize{(Fig.~\ref{fig:ten_cloth})}  \end{tabular}
& $11$ & $500\text{K}$ & $338\text{K}$ & $677\text{K}$ & \begin{tabular}{@{}c@{}} $28\text{K}$ \\ $7\text{M}$  \end{tabular} & $120|30$ & $\frac{1}{150}$ & $24.1$ & $8.4|9.1|0.3|8.8$ &$0.3$ &\begin{tabular}{@{}c@{}} $160$ \\ $3\cdot10^{-4}$  \end{tabular} &$1\cdot10^{-3}$  & \begin{tabular}{@{}c@{}} $35.9$ \\ ($31$)  \end{tabular} & $6.6$ & $16.6$ \\

\whline{0.5pt}
\begin{tabular}{@{}c@{}} \scriptsize{Just folding} \\ \scriptsize{(Fig.~\ref{fig:fold})}  \end{tabular}
& $5$ & $485\text{K}$ & $320\text{K}$ & $498\text{K}$ &\begin{tabular}{@{}c@{}} $23\text{K}$ \\ $7\text{M}$  \end{tabular} & $120|30$ & $\frac{1}{150}$ & $7.6$  & $5|7.5|0.5|7$ &$0.3$ & \begin{tabular}{@{}c@{}} $160$ \\ $5\cdot10^{-4}$  \end{tabular}&$5\cdot10^{-4}$  & \begin{tabular}{@{}c@{}} $64.4$ \\ ($46$)  \end{tabular} & $3.6$ & $12.4$ \\

\whline{0.5pt}
\begin{tabular}{@{}c@{}} \scriptsize{Cloth blender} \\ \scriptsize{(Fig.~\ref{fig:blend})}  \end{tabular}
& $8$ & $820\text{K}$ & $540\text{K}$ & $992\text{K}$ & \begin{tabular}{@{}c@{}} $160\text{K}$ \\ $26\text{M}$  \end{tabular} & $120|30$ & $\frac{1}{150}$ & $25.4$  & $10|11.6|2.6|9$ &$0.3$
  & \begin{tabular}{@{}c@{}} $160$ \\ $3\cdot10^{-4}$  \end{tabular} &$5\cdot10^{-4}$  & \begin{tabular}{@{}c@{}} $160.7$ \\ $(73)$  \end{tabular} & $10.2$ & $5.2$ \\
  
\whline{0.5pt}
\begin{tabular}{@{}c@{}} \scriptsize{Cover the ship} \\ \scriptsize{(Fig.~\ref{fig:cover_ship})}  \end{tabular}
& $5$ & $930\text{K}$ &941$\text{K}$ & $1.2\text{M}$ & \begin{tabular}{@{}c@{}} $63\text{K}$ \\ $9.2\text{M}$  \end{tabular} & $120|30$ & $\frac{1}{150}$ & $31.8$ & $8|8.3|1.2|7.1$ & $0.8|0.9$ &\begin{tabular}{@{}c@{}} $160|8\cdot 10^3$ \\ $2\cdot10^{-4}$  \end{tabular} &$5\cdot10^{-4}$  & \begin{tabular}{@{}c@{}} $117.8$ \\ $(51)$  \end{tabular} & $5.3$ &$7.2$ \\

\whline{0.5pt}
\begin{tabular}{@{}c@{}} \scriptsize{Funnel} \\ \scriptsize{(Fig.~\ref{fig:friction})}  \end{tabular}
& $6$ & $692\text{K}$ & $458\text{K}$ & $923\text{K}$ & \begin{tabular}{@{}c@{}} $276\text{K}$ \\ $32\text{M}$  \end{tabular} & $120|30$ & $\frac{1}{150}$ & $32.8$  & $17|13|2|9$ &$0.3$ & \begin{tabular}{@{}c@{}} $160$ \\ $3\cdot10^{-4}$  \end{tabular}&$5\cdot10^{-4}$  & \begin{tabular}{@{}c@{}} $112.4$ \\ $(52)$ \end{tabular} & $8.5$ & $6.6$ \\

\whline{0.5pt}
\begin{tabular}{@{}c@{}} \scriptsize{Kicking} \\ \scriptsize{(Fig.~\ref{fig:kicking})}  \end{tabular}
& $2$ & $450\text{K}$ & $294\text{K}$ & $525\text{K}$ & \begin{tabular}{@{}c@{}} $47\text{K}$ \\ $4\text{M}$  \end{tabular} & $120|30$ & $\frac{1}{200}$ & $11.6$  & $14|11|0.6|10$ &$0.3$ &\begin{tabular}{@{}c@{}} $200$ \\ $3\cdot10^{-4}$  \end{tabular} &$3\cdot10^{-4}$  & \begin{tabular}{@{}c@{}} $113.5$ \\ $(94)$  \end{tabular} &$8.5$&$6.8$ \\

\whline{0.5pt}
\begin{tabular}{@{}c@{}} \scriptsize{Fashion show} \\ \scriptsize{(Fig.~\ref{fig:teaser})}  \end{tabular}
& $2$ & $1.1\text{M}$ & $656\text{K}$ & $996\text{K}$ & \begin{tabular}{@{}c@{}} $84\text{K}$ \\ $10\text{M}$  \end{tabular} & $120|30$ & $\frac{1}{200}$ & $28.4$  & $14|14.7|1.6|13.1$ &$0.3$ &\begin{tabular}{@{}c@{}} $160$ \\ $3\cdot10^{-4}$  \end{tabular} &$5\cdot10^{-4}$  & \begin{tabular}{@{}c@{}} $167.3$ \\ $(67)$  \end{tabular}& $12.4$ &$4.8$ \\
\whline{1.15pt}
\end{tabular}
\end{center}
}
\end{table*}

\paragraph{Our method vs PD-BFGS}
Another relevant competitor is from a recent contribution by \citet{li2023subspace} or PD-BFGS. PD-BFGS designs a two-step global solve combining Jacobi iteration with BFGS method in a spline-based subspace. PD-BFGS is also based on DBB, and thus slower than our method for IPC-based collision processing. At the global stage, PD-BFGS employs a reduced quasi-Newton procedure in a B-spline subspace. This subspace is of high dimension i.e., $r \approx 10,000$. As a result, PD-BFGS is much slower than the pre-computed subspace update used in our simulation. A concrete experiment is reported in Fig.~\ref{fig:knot}, where we pull two intertwining cloth strips to make a tight knot. There are $104$K vertices on the strips. Our method handles this example at an interactive rate of $11.8$ FPS, while PD-BFGS needs seconds for each frame. 

\begin{figure}
  \centering
  \includegraphics[width=\linewidth]{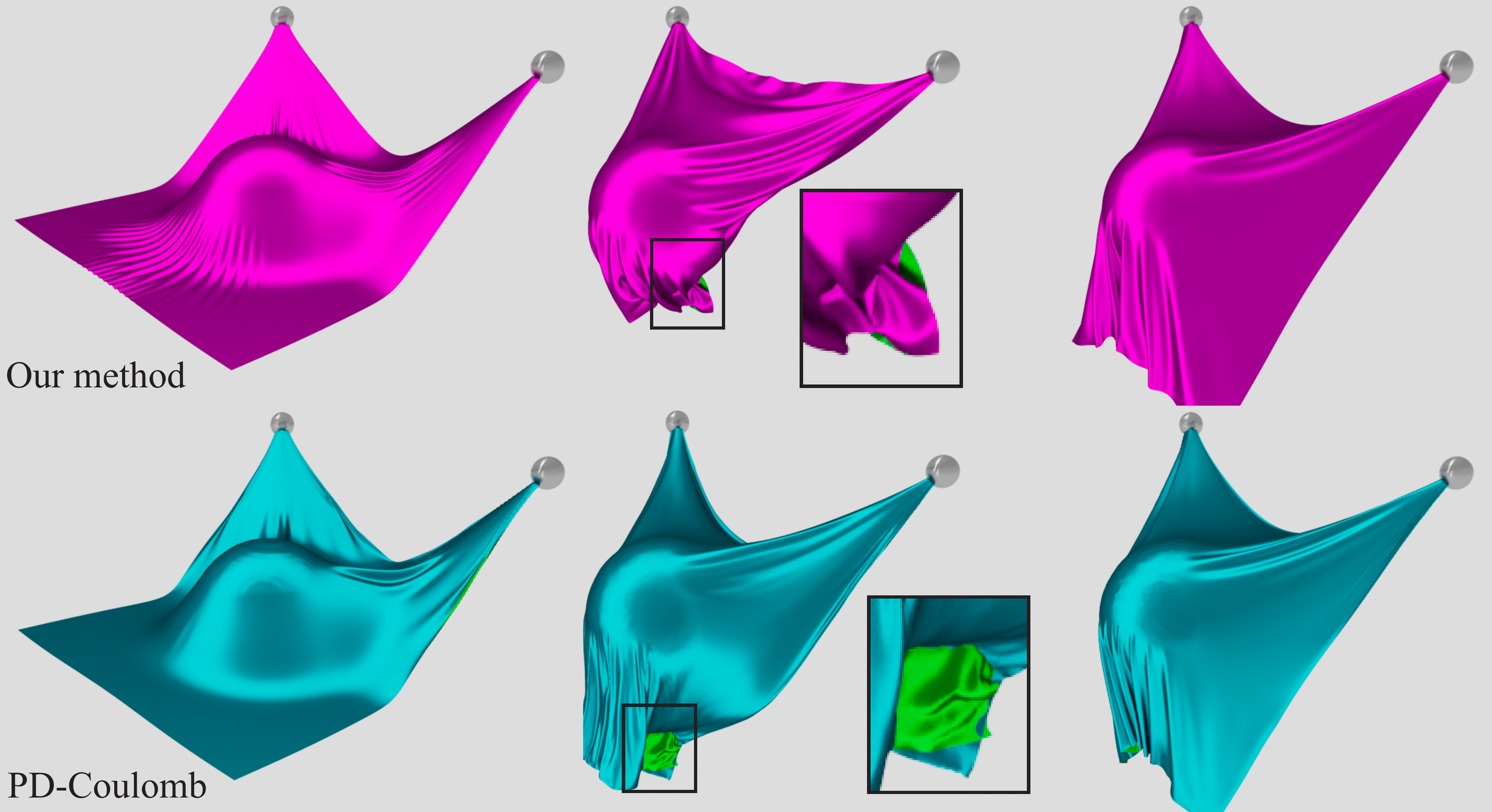}
  \caption{\textbf{Drape on a sphere.}~~A tablecloth of $40$K vertices drapes on the sphere. We compare our method with PD-Coulomb~\cite{ly2020projective}. PD-Coulomb models Coulomb friction, while our method uses a lagged friction Hessian at each LG iteration as in~\cite{li2023subspace}.}
  \label{fig:pd_coulomb}
\end{figure}

\begin{figure}
  \centering
  \includegraphics[width=\linewidth]{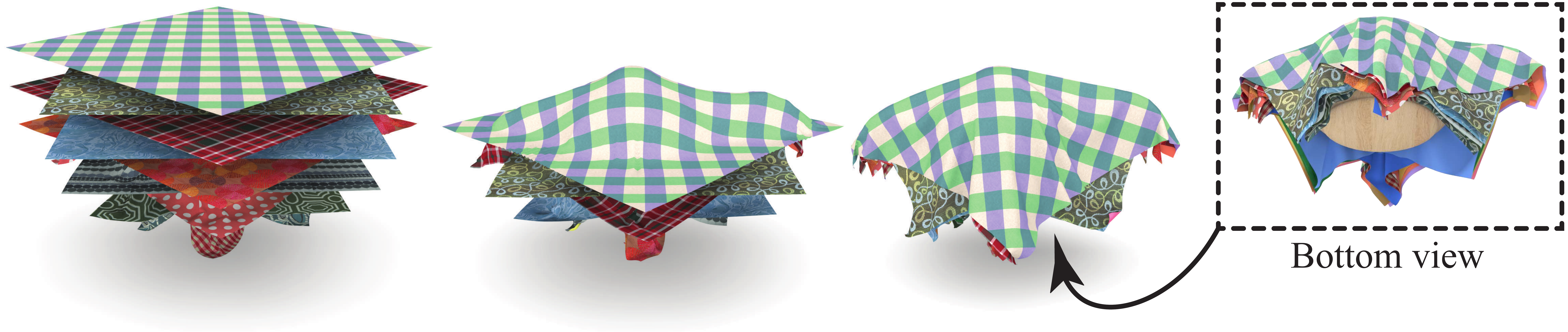}
  \caption{\textbf{Stack on the teapot.}~~We stack ten pieces of tablecloth on the teapot simultaneously. The free fall of stacking clothes generates a large number of intertwining cloth-cloth collisions. EXP and partial CCD robustly handle this challenging scene and produce high-quality animation.}
  \label{fig:ten_cloth}
\end{figure}
\begin{figure}
  \centering
  \includegraphics[width=\linewidth]{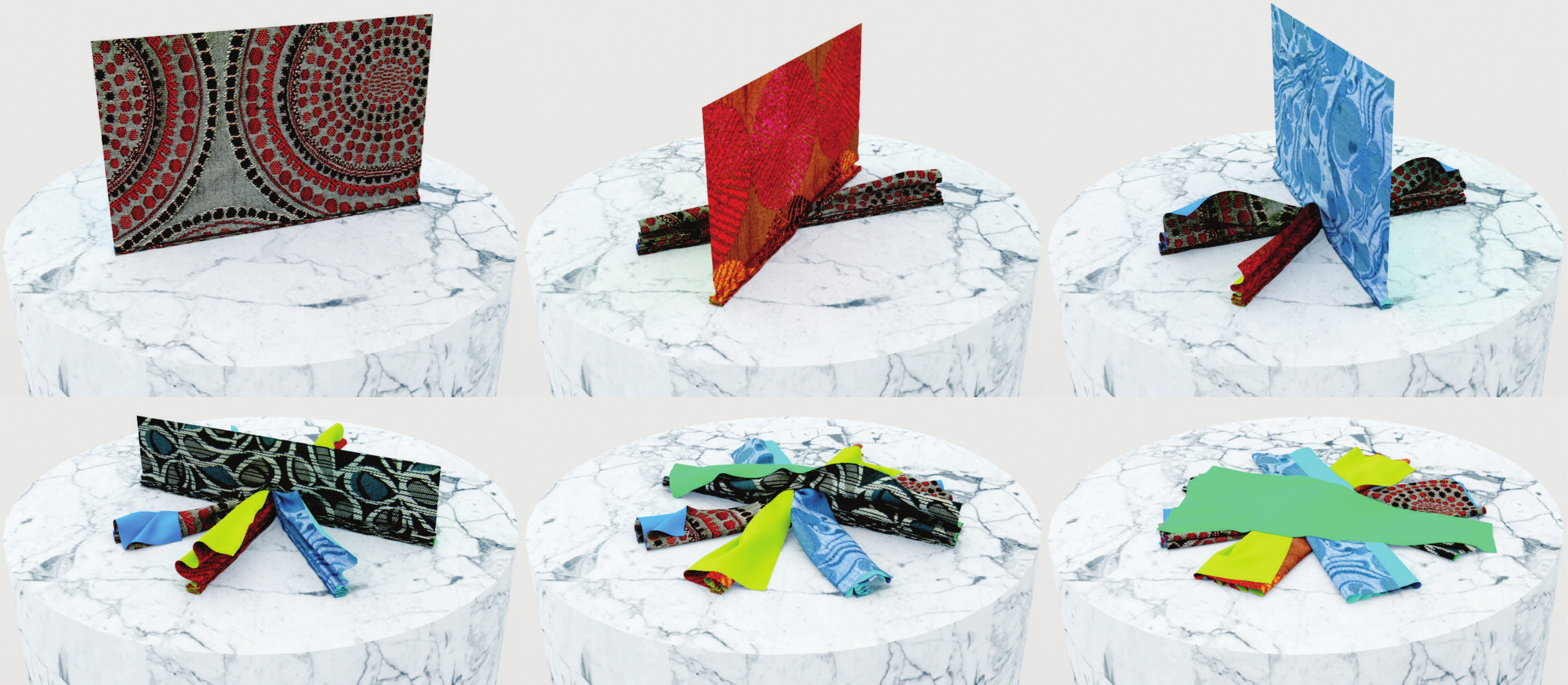}
  \caption{\textbf{Just folding.}~~In this example, four clothes fall vertically on the desk one by one with different orientations. The contact between the cloth and the desktop folds the cloth with complex self-collisions. This is a good stress test because the velocity of falling cloth is high. The resulting multi-layer self-collisions are particularly challenging. There are nearly $500$K DOFs involved, and the simulation runs at $12$ FPS.}
  \label{fig:fold}
\end{figure}

\paragraph{Our method vs PD-Coulomb}
Our method is compatible with various friction models. In our implementation, we follow the strategy used in~\cite{li2023subspace}, which leverages a quadratic barrier proxy to estimate the Hessian of the friction energy to simulate frictional surfaces with different friction coefficients -- in a way similar to the original IPC. Alternatively, \citet{ly2020projective} (PD-Coulomb) show that it is possible to incorporate full Coulomb friction by a novel splitting strategy so that the global matrix is kept constant. To this end, we show a side-by-side comparison between our method and PD-Coulomb~\cite{ly2020projective}.

\begin{figure*}
  \centering
  \includegraphics[width=\linewidth]{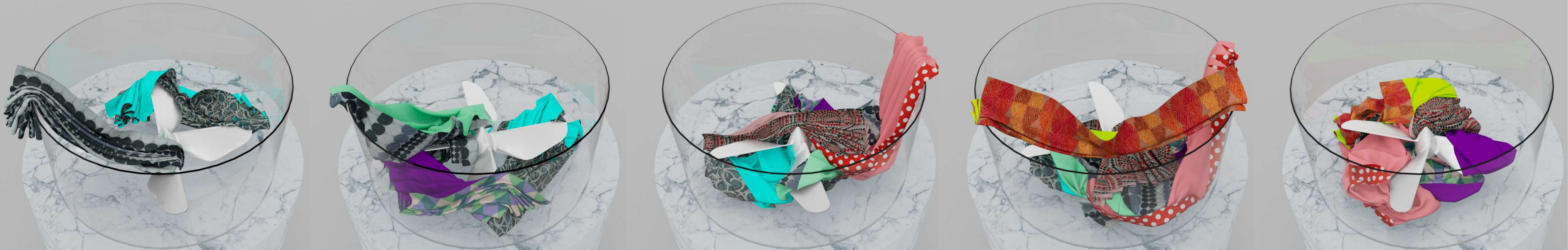}
  \caption{\textbf{Cloth blender.}~~We drop six clothes into a bowl-like container. The blender rotates and stirs all the clothes. This example generates fast rotational collisions between clothes and the collider, as well as a large number of cloth-cloth collisions. The simulation has $820$K DOFs and runs at about $5$ FPS.}
  \label{fig:blend}
\end{figure*}

\begin{figure*}
  \centering
  \includegraphics[width=\linewidth]{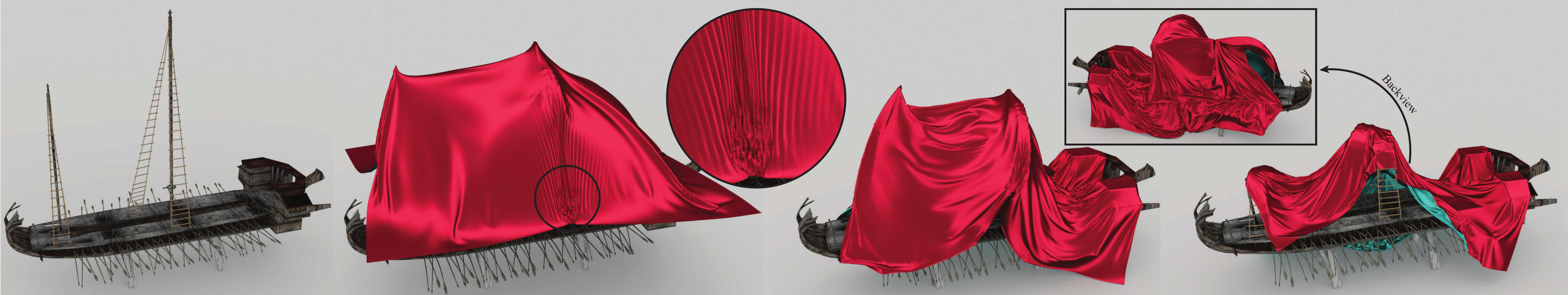}
  \caption{\textbf{Cover the ship.}~~We cover a deformable barbarian ship with a piece of heavy cloth ($\rho_m =  0.8 kg/m^2$). The soft masts and ladders on the ship bend under the weight of the cloth cover. The ship-cloth contacts generate fine wrinkles that delineate the shape features of the ship. There are $540$K DOFs on the cloth and $390$K DOFs on the barbarian ship. With {$\Delta t = 1/150$}, our method resolves all the cloth-cloth, cloth-ship, and ship-ship collisions and runs at $7.2$ FPS.}
  \label{fig:cover_ship}
\end{figure*}
In this experiment, we drape a piece of square cloth of $40$K vertices on a sphere. The size of the cloth is $1m \times 1m$, and the time step is $\Delta t = 1/200$ sec. It can be seen from Fig.~\ref{fig:pd_coulomb} (and also see the supplementary video) that both methods capture the frictional contact between the cloth and the sphere and produce similar animation results. In fact, the splitting method proposed in~\cite{ly2020projective} can also be integrated with our pipeline. PD-Coulomb adopts DCD, and we follow the default setting as in the published code of~\cite{ly2020projective}. The collision tolerance is set as $1mm$. This means as long as the distance between two triangles is smaller than $1mm$, DCD will generate a contact. While this is a conservative configuration, considering the movement of the cloth is moderate, one can still spot minor inter-penetration among the cloth triangles. On the other hand, our method uses CCD and a line search filtering at the end of each step, which guarantees all the triangles are separate. 

It should be mentioned that PD-Coulomb is CPU-based, using \texttt{OpenMP} to speed up pre-factorized global solve, while our method is fully GPU-based. As a result, our method runs faster (at $25$ FPS) than PD-Coulomb by 30 times. Such performance difference largely comes from the hardware platform. Nevertheless, our method is orthogonal to PD-Coulomb as we do not focus on friction modeling but more on solver optimization and collision detection.

\begin{figure}
  \centering
  \includegraphics[width=\linewidth]{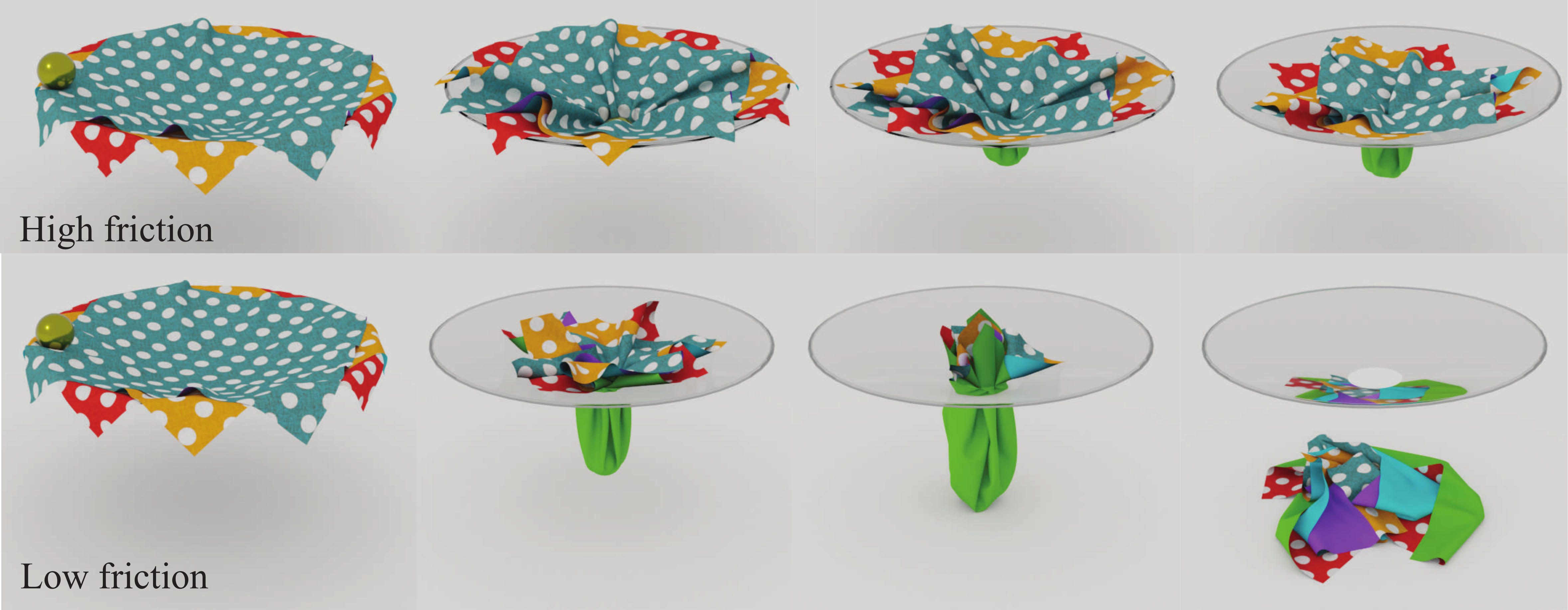}
  \caption{\textbf{Funnel.}~~We place a rigid and heavy ball (with affine body dynamics~\cite{lan2022abd}) on a funnel covered by three layers of cloth. The cloths hold the sphere under high frictional contacts (top). When the friction is not strong enough, all the cloths eventually fall on the ground (bottom). The simulation involves nearly $700$K DOFs, and our method runs at $6.6$ FPS.}
  \label{fig:friction}
\end{figure}

\begin{figure*}
  \centering
  \includegraphics[width=\linewidth]{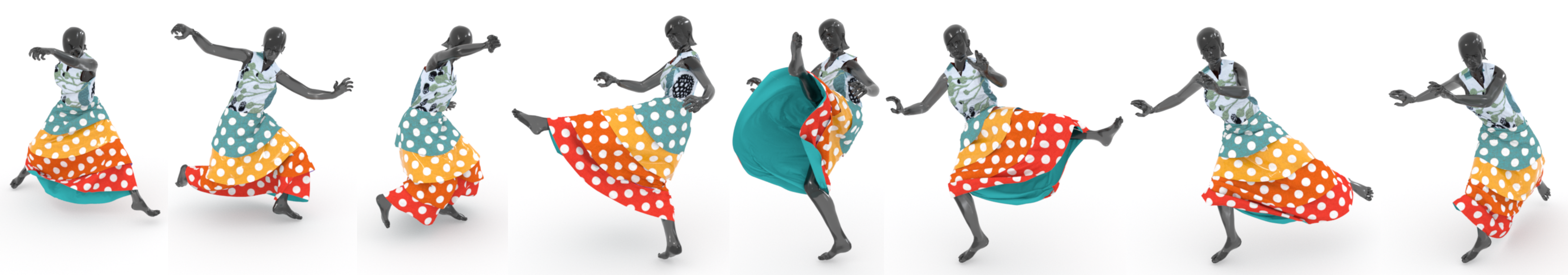}
  \caption{\textbf{Kicking.}~~In this example, the virtual character quickly performs a kicking action, which leads to nonlinear animation effects on the multi-layer skirt. Our method produces high-quality results. The frame rate reaches $6.8$ FPS, which is two-order faster than the state-of-the-art GPU simulation~\cite{li2023subspace}.}
  \label{fig:kicking}
\end{figure*}

\subsection{More experiments}
We have tested our method in a wide range of simulation scenes. The detailed time statistics of all the experiments shown in the paper are reported in Tab.~\ref{tab:time}. Figs.~\ref{fig:ten_cloth}, \ref{fig:fold}, and \ref{fig:blend} show three tests with massive collisions. In Fig.~\ref{fig:ten_cloth}, we test the robustness of our method under extensive stacking. In this test, ten falling tablecloths hang on the teapot. This simulation generates a lot of overlapping collisions when clothes stack under gravity. Fig.~\ref{fig:fold} reports another test where we fold four clothes by vertically dropping them on the desk. As the cloth makes contact with the desktop, it folds in a zig-zag pattern, resulting in a large number of self-collisions, particularly edge-edge collisions. This experiment is a good stress test showing the robustness of a cloth simulator. Fig.~\ref{fig:blend} mimics a washer with six clothes in a bowl-shaped container. The scene also involves a large number of dynamic collisions. Our method produces interesting animations in all of those experiments, and the results are free of any inter-penetration.

Our method is also capable of simulating deformable objects. In addition to the example reported in Fig.~\ref{fig:animal}, we give another experiment with two-way coupling between cloth and deformable body under complex collisions. As shown in Fig.~\ref{fig:cover_ship}, a heavy cloth falls onto a barbarian ship. The ship base is fixed, and it has soft masts and ladders. As the cloth falls down, we can see detailed wrinkles at the contacting area between the cloth and the ship. There are $540$K DOFs on the cloth and $390$K DOFs on the ship. Our method generates interesting animation results at 7.2 FPS.    

Fig.~\ref{fig:friction} shows two more simulations under different frictional setups. Three pieces of cloth cover a funnel, and we place a rigid/heavy ball on the funnel. We use affine body dynamics~\cite{lan2022abd} to simulate the ball's motion. There are nearly $700$K DOFs in this scene, and our method produces plausible dynamics under different friction settings. The simulation runs at $6.6$ FPS. As discussed, we can switch to a more accurate frictional model either as in~\cite{ly2020projective}, which is based on complementarity programming or as in~\cite{li2020incremental}, which is based on the interior-point method.

Cloth animations play a pivotal role in the realm of digital fashion and design. The proposed method enhances this aspect by enabling virtual characters to interact seamlessly with a variety of garments, yielding high-quality simulations. As a demonstration of its capabilities, we report two additional examples. The first features a character executing a kicking action, illustrating the dynamic interaction between the motion and the garment (Fig.~\ref{fig:kicking}). This is a numerically challenging test as the body undergoes swift movements. Most DCD-based methods are unable to capture such high-velocity garment dynamics. The second example showcases a virtual model on a fashion runway as shown in Fig.~\ref{fig:teaser}. This model, attired in a knee-length skirt, walks to the front of the stage before turning around to return. Throughout this sequence, our method meticulously captures the intricacies of the motion. Please refer to the supplementary for the animated results, where we report two sequences of animations

\section{Conclusion \& Limitation}\label{sec:conclusion}
This paper presents a parallelizable cloth simulation framework, that delivers high-quality animation results, keeps computations lightweight, and separates all the triangles on the cloth models. To the best of our knowledge, such combined efficiency, quality, and performance are not possible with existing cloth animation algorithms. To achieve this goal, we employ a non-distance barrier, which is both simple and effective. This new barrier model allows the simulation to skip the computation of the distance between primitive pairs, and the weight of each collision constraint becomes self-adjusting during LG iterations given the current active set. The subspace reuse scheme significantly pushes the performance of the solver with minimum costs. By observing the fact that the low-frequency subspace is less sensitive to high-frequency collisional deformations, we reuse the rest-shape modal global matrix to solve global cloth deformation. The subspace matrix update is also efficient and can be pre-computed. The residual forwarding helps mitigate the dumpling artifacts due to small-TOI line searches. We show that this approach yields visually plausible animations with the penetration-free guarantee and makes the simulation efficient for high-resolution scenes. 

\begin{table}
\caption{\textbf{More statistics on parameters.} We report total LG iterations and time cost (in ms) using different simulation parameters. Detailed results (including artifacts and simulation failure cases) are presented in the supplementary videos.}\label{tab: lg_limitation}
{\scriptsize \fontfamily{ppl}\selectfont 
\begin{center}
\begin{tabular}{p{0.5cm}||c|c|c|c}
\whline{1.15pt}
\multicolumn{5}{c}{Cloth folding}\\
\whline{0.75pt}
${\Delta t} (\text{s})$ & $\|\Delta x\|=1\cdot 10^{-4}$ & $\|\Delta x\|=5\cdot 10^{-4}$ & $\|\Delta x\|=1\cdot 10^{-3}$ & $\|\Delta x\|=1\cdot 10^{-2}$\\
\whline{0.75pt}
$1/150$ & $39 (16.3\text{ms})$ & $18 (8.3\text{ms})$ & $15 (6.9\text{ms})\;$ \textbf{ Jittery} & $4 (2.8\text{ms})$ \textbf{Jittery} \\
\whline{0.75pt}
$1/100$ & $55 (24.7\text{ms})$ & $26 (13.5\text{ms})$ &$20 (11.1\text{ms})$  \textbf{Jittery} & \textbf{Fail}\\
\whline{0.75pt}
$1/50$ & $95 (56.4\text{ms})$ & $45 (26.6\text{ms})$ & $40 (20.9\text{ms})$  \textbf{Jittery} & \textbf{Fail}\\
\whline{1.15pt}
\multicolumn{5}{c}{Cloth twisting}\\
\whline{0.75pt}
${\Delta t} (\text{s})$ & $\|\Delta x\|=1\cdot 10^{-4}$ & $\|\Delta x\|=5\cdot 10^{-4}$ & $\|\Delta x\|=1\cdot 10^{-3}$ & $\|\Delta x\|=1\cdot 10^{-2}$\\
\whline{0.75pt}
$1/150$ & $214 (64.8\text{ms})$ & $171 (52.2\text{ms})$ & $157 (47.4\text{ms})$ & $72 (21.6\text{ms})$ \\
\whline{0.75pt}
$1/100$ & $229 (69.7\text{ms})$ & $182 (56.1\text{ms})$ &$163 (49.4\text{ms})$ & \textbf{Fail}\\
\whline{0.75pt}
$1/50$ & $255 (77.3\text{ms})$ & $202 (61.1\text{ms})$ & $171 (51.9\text{ms})$ & \textbf{Fail}\\
\whline{1.15pt}
\end{tabular}
\end{center}
}
\end{table}

\begin{figure}
  \centering
  \includegraphics[width=0.9\linewidth]{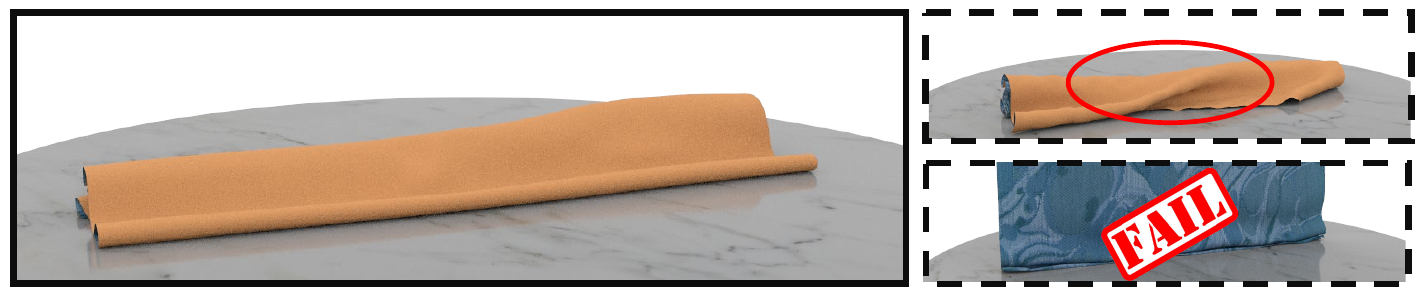}
  \includegraphics[width=0.9\linewidth]{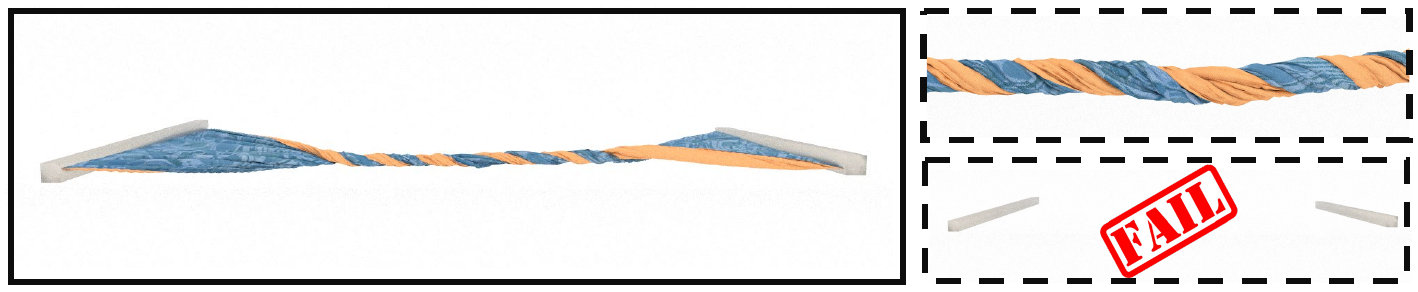}
  \caption{\textbf{Folding and twisting.}~~As a non-Newton method, the performance of our method is correlated with time step size and termination criteria. RF may introduce unnatural jittery artifacts and may cause simulation failure when the residual error is too large. The animated results are available on supplementary video.}
  \label{fig:limitation}
\end{figure}

Our method has some limitations. First, our method is based on projective dynamics. Therefore, the elasticity model ($\Psi$) should be shaped in a quadratic form. This prevents us from incorporating more complex fabric models in the simulation e.g., homogenized~\cite{sperl2020homogenized} or data-driven models~\cite{feng2022learning}. However, we believe it is possible to combine our subspace reuse strategy with block descent methods~\cite{lan2023second} for general materials. 
Residual forwarding trick becomes less effective if the residual errors are too large and may cause simulation failure: in some situations, RF can introduce unnatural bumpy artifacts, which may be even more visually disturbing (see Tab.~\ref{tab: lg_limitation}, two columns on the right, Fig.~\ref{fig:limitation} and supplementary videos). While our method adapts to various simulation parameters, it is more sensitive to timestep size and termination criteria compared to the projection-Newton.


\begin{acks}
We thank reviewers for their detailed and constructive comments. Yin Yang is partially supported by NSF under grant numbers 2301040, 2008915, 2244651, 2008564. Chenfanfu Jiang is supported in part by NSF CAREER 2153851, CCF 2153863, ECCS-2023780.
\end{acks}

\bibliographystyle{ACM-Reference-Format}
\bibliography{ref}

\appendix

\end{document}